\documentclass[journal]{IEEEtran}

\usepackage{amsmath,amsfonts}
\usepackage{algorithmic}
\usepackage{array}
\usepackage[caption=false,font=normalsize,labelfont=sf,textfont=sf]{subfig}
\usepackage{textcomp}
\usepackage{stfloats}
\usepackage{url}
\usepackage{verbatim}
\usepackage{graphicx}
\def\BibTeX{{\rm B\kern-.05em{\sc i\kern-.025em b}\kern-.08em
    T\kern-.1667em\lower.7ex\hbox{E}\kern-.125emX}}
\usepackage{balance}

\usepackage{amsthm}
\usepackage{xcolor}
\usepackage{hyperref}
\hypersetup{colorlinks=true,
    linkcolor=blue,
    citecolor=blue,
    urlcolor=blue}
\usepackage[capitalise]{cleveref}
\usepackage{cite}
\usepackage{algorithm}
\usepackage{mathtools}
\usepackage{bbold} 
\usepackage{multirow}

\crefname{appendix}{Appendix}{Appendices}
\Crefname{appendix}{Appendix}{Appendices}

\crefformat{equation}{(#2#1#3)}
\crefrangeformat{equation}{(#3#1#4) to~(#5#2#6)}
\crefmultiformat{equation}
  {(#2#1#3)}
  { and~(#2#1#3)}
  {, (#2#1#3)}
  { and~(#2#1#3)}

\newcommand{\id}{\mathbb{1}}
\newcommand{\tr}{\operatorname{Tr}}
\newcommand{\srf}{\text{SRF}}
\newcommand{\smallt}{\varepsilon_t}
\newcommand{\diag}{\operatorname{diag}}
\newcommand{\var}{\operatorname{var}}
\newcommand{\cov}{\operatorname{cov}}

\newcommand{\rank}{\operatorname{rank}}
\newcommand{\spec}{\operatorname{spec}}
\renewcommand{\Im}{\operatorname{Im}}
\renewcommand{\Re}{\operatorname{Re}}

\theoremstyle{plain}
\newtheorem{theorem}{Theorem}
\theoremstyle{plain}
\newtheorem{corollary}[theorem]{Corollary}
\theoremstyle{plain}
\newtheorem{lemma}{Lemma}
\theoremstyle{definition}

\theoremstyle{plain}

\theoremstyle{plain}

\AddToHook{env/corollary/begin}{\crefalias{theorem}{corollary}}

\crefname{corollary}{Corollary}{Corollaries}
\Crefname{corollary}{Corollary}{Corollaries}

\definecolor{jonas}{rgb}{0.9,0.17,0.31}

\definecolor{ingo}{rgb}{0.9,0.8,0.11}

\begin{document}

\title{Estimating the generator of \\ matrix exponential time-series}

\author{Jon{\' a}{\v s} Fuksa, Jens Eisert, and 
Ingo Roth
\thanks{J. Fuksa and J. Eisert are with the 
Dahlem Center for Complex Quantum Systems, Freie Universit\"at Berlin, 14195 Berlin, Germany (e-mail: jonas.fuksa@fu-berlin.de).}
\thanks{J. Eisert is with the Helmholtz-Zentrum Berlin f{\"u}r Materialien und Energie, Berlin, Germany.}
\thanks{I. Roth is with the Quantum Research Center, Technology Innovation Institute (TII), Abu Dhabi, United Arab Emirates.}
}

\maketitle

\begin{abstract}
    The achievable precision in estimating the spectrum of a time-series signal that is promised to be sparse in Fourier space deteriorates rapidly as the spectral peaks get closer together, as soon as we introduce noise.
    Here, we study the related task of generator estimation: estimating the generator of a matrix exponential time-series and its eigenvalues.
    We obtain rigorous non-asymptotic precision guarantees on genESPRIT, a practical algorithm solving generator estimation, and show that this task can be solved with precision independent of the spectral spacing.
    The guarantees are complemented by a Cramér--Rao analysis of a relaxation of the problem.
    In particular, in the case of unitary data with bounded additive noise with effective entrywise variance $\sigma^2$, we show that genESPRIT estimates the associated propagator matrix in Frobenius norm with precision scaling as $\tilde{\mathcal{O}}(\sigma NL^{-\frac12})$ in the length of the time-series $L$ and matrix dimension $N$, matching the corresponding spectrum-independent Cramér--Rao standard deviation scaling up to logarithmic factors.
    We prove that the matching distance error in the spectrum estimated by genESPRIT scales as $\tilde{\mathcal{O}} (\sigma^2 N^3 L^{-1} + \sigma N^{\frac32} L^{-\frac32} )$, which we compare to the lower bound on the corresponding Cramér--Rao bound with the asymptotic standard deviation scaling $\mathcal{O}(\sigma L^{-\frac32})$.
    These guarantees provide the first rigorous bounds on solving the generator estimation problem.
    As the matrix exponential time-series can be obtained from linearly independent vector time-series by concatenation, this work opens up a new direction in spectral estimation, which circumvents the problem of crowded spectra, under the additional assumption of linear independence.
\end{abstract}

\begin{IEEEkeywords}
    Signal processing, non-asymptotic performance analysis, ESPRIT, matrix perturbation theory, super-resolution, matrix time-series, genESPRIT.
\end{IEEEkeywords}

\section{Introduction}\label{sec:introduction}

\IEEEPARstart{T}{he} task of estimating spectral properties of a signal from noisy samples is fundamental in signal processing.
Its applications range from imaging to direction-of-arrival estimation and identification of dynamical systems~\cite{stoicaMaximumLikelihoodMethods1990,stoicaSpectralAnalysisSignals2005,isermannIdentificationDynamicSystems2011,schmidtSignalSubspaceApproach1981a,slepianEstimationSignalParameters1954,gariniSpectralImagingPrinciples2006}.
General-purpose methods based on the \emph{discrete Fourier transform} (DFT) are fundamentally limited in resoltuion by the Rayleigh limit: spectral peaks can be resolved if they are separated by at least $2\pi/L$, where $L$ is the extent of the signal.

However, advances in the field of \emph{compressed sensing} have shown that the achievable precision of an estimation procedure often depends strongly on the assumptions about the signal~\cite{donohoCompressedSensing2006,chenSpectralCompressedSensing2013}, raising the question of how well spectral properties can be estimated when some information is available in advance about the incoming signal.
Independently, in the field of imaging, \emph{super-resolution} (SR) techniques circumventing the $1/L$ limit have been developed.
While some of these techniques achieve SR by purely optical phenomena, such as evanescent waves~\cite{lindbergMathematicalConceptsOptical2012,puschmannSuperresolutionAstronomicalImaging2005,milanfarSuperResolutionImaging2017,greenspanSuperResolutionMedicalImaging2009,schermellehSuperresolutionMicroscopyDemystified2019}, a plethora of SR techniques turned out to be akin to compressed sensing: they achieve SR by placing structural assumptions on the signal~\cite{candesMathematicalTheorySuperresolution2014a,tangCompressedSensingGrid2013,mishraSpectralSuperResolutionPrior2015,donohoSuperresolutionSparsityConstraints1992}.
Assuming spectral sparsity has proven particularly useful in this context.

While most SR techniques are designed for scalar or vector data, in many applications we have access to data in the form of a \emph{matrix exponential} time-series $y(t) = A e^{i\Omega t} B \in \mathbb{C}^{N \times N}$ or, more realistically, its noisy samples
\begin{equation}\label{eq:intro matrix exponential time-series}
    y[\ell] = A e^{i\ell\Omega \smallt} B + \eta[\ell], \quad \ell \in \{0, \dots, L-1\},
\end{equation}
where $\smallt>0$ is a time step, $\Omega = \diag(\omega), \ \omega \in \mathbb{R}^{N}$, $\eta[\ell]$ is additive noise and $A, B \in \mathbb{C}^{N \times N}$ are invertible.
For instance, this is the case when the dynamics of a Koopman dynamical system is probed with linearly independent initial conditions, in DOA estimation with a rectangular grid antenna array, in time evolution of closed quantum systems~\cite{hangleiterRobustlyLearningHamiltonian2024a} or when measuring correlation functions of quantum fields in a continuous matrix product state~\cite{steffensExperimentalQuantumfieldTomography2015,steffensQuantumFieldTomography2014}. 
In the case when only vector time-series data $\vec{x}(t) = A e^{i\Omega t}\vec{x}_0$ are available, we can obtain linearly independent columns by varying the initial vector $\vec{x}_0$ or, for non-degenerate subspaces of the generator, by using shifted data $\vec{x}'(t) = \vec{x}(t + \tau)$ for various delays $\tau$.

The central focus of this work is the problem we coin \emph{generator estimation}.
Here, the task is to estimate $G \coloneqq A \Omega A^{-1} \in \mathbb{C}^{N\times N}$ and $\Omega \in \diag{\mathbb{R}^N}$ from data $y \in \mathbb{C}^{N \times N \times L}$ given by~\cref{eq:intro matrix exponential time-series} with independent additive noise $\eta$.
The quality of the estimates $\hat G$ and $\hat \Omega$ is measured by the propagator Frobenius norm error $\|e^{i \smallt \hat G} - e^{i \smallt G}\|_F$ and the matching distance on the circle $\|\hat \Omega - \Omega\|_{\text{md}, \mathbb{T}}$ respectively.
To measure the error in the generator $\hat G$ itself, assumptions need to be placed on the norm of $G$ to prevent aliasing issues and to fix the correct branch of the matrix logarithm.

The signal $y(t)$ is an $N\times N$ complex matrix parametrised by time.
Each entry is a scalar time-series with the Fourier transform
\begin{equation}
    \mathcal{F}[y_{i,j}](\omega) = \sum_{k=0}^{N-1} A_{i,k}B_{k,j} \delta(\omega - \omega_k),
\end{equation}
i.e., a signal with a sparse spectrum with at most $N$ peaks, which can be turned on and off by varying $i,j$, but their location stays fixed.
The problem of generator estimation can therefore 
be thought of as the \emph{simultaneous} estimation of the spectral properties of $N^2$ scalar time-series signals with Fourier transforms supported within a common set of at most $N$ peaks.

If each entry of the signal is analysed independently from the others, the precision limits are well understood by super-resolution theory, going back over fifty years to the work of Rife and Boorstyn~\cite{rifeSingleToneParameter1974, rifeMultipleToneParameter1976}.
They found that the best achievable asymptotic precision in estimating the frequency of a single tone scales as $L^{-\frac32}$ and that  multi-tone estimation depends strongly on the minimal frequency spacing $\Delta$. 
For $\Delta > \frac{2\pi}{L}$ the limiting asymptotic precision of the multi-tone case reduces to the single-tone case, while when $\Delta < \frac{2\pi}{L}$ the precision deteriorates rapidly.
Subsequent rigorous non-asymptotic analyses~\cite{donohoSuperresolutionSparsityConstraints1992,demanetRecoverabilityLimitSuperresolution2015,liStableSuperresolutionLimit2021,batenkovSuperresolutionNearcollidingPoint2021} introduce the \emph{super-resolution factor} $\srf \coloneqq \frac{2\pi}{\Delta L}$, which marks the phase transition between the \emph{well-separated} regime $\srf<1$ and the \emph{clustered} regime $\srf>1$.  
In the well-separated regime the problem behaves like estimating a single tone, in the clustered regime the best achievable precision of estimating $N$ tones scales as $\srf^{2N-1}$.

In tandem with exploring the fundamental limits of multi-tone estimation, a range of practical algorithms that provably solve this problem with optimal or near-optimal precision has been proposed.
One line of research, close in spirit to compressed sensing techniques, tackles the problem by finding the best-fitting sparse frequency vector~\cite{liApproximateSupportRecovery2020,tangCompressedSensingGrid2013,mishraSpectralSuperResolutionPrior2015,candesSuperResolutionNoisyData2013,liApproximateSupportRecovery2020,liElementary$L^infty$Error2017,yangEnhancingSparsityResolution2016,duvalExactSupportRecovery2015,tangMinimaxLineSpectral2015,azaisSpikeDetectionInaccurate2015,batenkovStabilitySuperresolutionGeneralized2018,fernandez-grandaSuperresolutionPointSources2016,fernandez-grandaSupportDetectionSuperresolution2013,candesMathematicalTheorySuperresolution2014a}. 
By means of convex relaxation, these techniques find an approximately optimal solution by minimising the atomic norm of the spectral function.
Another line of research, pioneered over two hundred years ago by Prony~\cite{prony1795,pottsParameterEstimationExponential2010,pottsParameterEstimationNonincreasing2013}, has developed into a rich field of \emph{signal-subspace} based methods.
Modern examples include the MUSIC~\cite{schmidtSignalSubspaceApproach1981a,poratAnalysisAsymptoticRelative1988,liaoMUSICJointFrequency2014,liaoMUSICSinglesnapshotSpectral2016,stoicaMUSICMaximumLikelihood1989,liStabilitySuperresolutionMUSIC2022,kavehStatisticalPerformanceMUSIC1986} and ESPRIT~\cite{paulrajSubspaceRotationApproach1986,royESPRITestimationSignalParameters1989,fannjiangCompressiveSpectralEstimation2016,swindlehurstMultipleInvarianceESPRIT1992,yangNonasymptoticPerformanceAnalysis2023,badeauPerformanceESPRITEstimating2008,liSuperResolutionLimitESPRIT2020,linTimeFrequencyMultiInvarianceESPRIT2016} algorithms.

In the work~\cite{hangleiterRobustlyLearningHamiltonian2024a} the present authors have extended signal subspace techniques to the case of matrix exponential time-series, proposing the genESPRIT algorithm.
In this work, we rigorously analyse the non-asymptotic performance of genESPRIT in the case of unitary $A$ and $B$ and verify the results numerically.
By also studying the associated Cramér--Rao bounds on a relaxed version of the problem, we show that analysing all entries of the matrix time-series~\cref{eq:intro matrix exponential time-series} \emph{simultaneously} allows us to circumvent the phase transition at $\srf=1$ entirely.
The guarantees for genESPRIT are completely independent of the frequency spacing and the performance does not deteriorate even for completely degenerate spectra.
The dependence on $\srf$ is hence completely removed and the need for spectral separation is replaced by an orthonormality condition that $A$ and $B$ are unitary.
Numerically, we show that genESPRIT also works in the case of invertible $A, B$, showing that the weaker linear independence condition suffices.

The structure of the remainder of this work is as follows.
In~\cref{sec:cramer-rao} we analyse the asymptotic precision limits of the generator estimation problem by analysing the corresponding Cramér--Rao bound. 
Then, in~\cref{sec:genESPRIT} we describe genESPRIT, before turning to proving rigorous non-asymptotic guarantees on its performance in~\cref{sec:guarantees}.
We numerically demonstrate the performance of genESPRIT in~\cref{sec:numerics}.

\section{Notation}\label{sec:notation}
\noindent
In this section, we specify the notation that is made use of in the remainder of this work.
For an integer $k$, we label with $[k]$ the set $\{0, \dots, k-1\}$.
For $z \in \mathbb{C}$ we write $\arg(z)$ for its complex argument and $\Re(z)$ ($\Im(z)$) for its real (imaginary) part.
For any matrix $A$, we label with $A^T$ the matrix transpose, with $A^\dag$ the conjugate transpose and with $A^+$ the Moore-Penrose pseudoinverse.
We let $\kappa(A)$ be the condition number of $A$, $\spec(A)$ its spectrum and $\sigma(A)$ its singular values in decreasing order.
We label $\|A \|_\infty = \max \sigma(A)$ the operator norm and $\| A \|_F = \|\sigma(A)\|_2$ the Frobenius norm of a matrix, where $\|v\|_2$ for a vector $v$ is the Euclidean norm. 
For $N\times N$ matrices $A$ and $B$, these norms satisfy $\|AB\|_F \le \|A\|_\infty \|B\|_F$ and $\|A\|_\infty \le \|A\|_F \le \sqrt{N} \|A\|_\infty$ and they are subadditive and submultiplicative.
For a vector $v \in \mathbb{C}^N$ the diagonal matrix with $v$ on the diagonal is labeled by $D = \diag(v)$, similarly we write $v = \diag(D)$.
For $\hat \omega, \omega \in (-\pi, \pi]$ and $\hat\Omega = \diag(\hat \omega)$, $\Omega = \diag(\omega)$, the matching distance on the circle is defined by
\begin{equation}
    \|\hat \Omega - \Omega\|_{\text{md},\mathbb{T}} \coloneqq \min_{\pi\in \mathcal{S}_N} \max_k \|\hat\omega_k - \omega_{\pi(k)}\|_\mathbb{T},
\end{equation}
where $\|\cdot\|_\mathbb{T}$ is the metric on the circle $(-\pi, \pi]$ and $\mathcal{S}_N$ the symmetric group over $N$ elements.
We will also use the ordinary matching distance $\|\hat\Lambda - \Lambda\|_\text{md} \coloneqq \min_{\pi \in \mathcal{S}_N} \max_k |\hat \lambda_k - \lambda_{\pi(k)}|$.

\section{Lower bounds on generator estimation}\label{sec:cramer-rao}

\noindent
We will now discuss the theoretical limits of the generator estimation problem.
Note that in the definition we have made the choice to seek $A \Omega A^{-1}$ instead of $B^{-1} \Omega B$.
This is inconsequential as using the transpose of the data $y^T$ seeks the latter, up to a transpose.
Note that the choice of measuring the quality of the estimator using $\|e^{i \smallt \hat G} - e^{i \smallt G}\|_F$ means that generator estimation will not have a unique solution and $\hat G' = \hat G + 2\pi \smallt^{-1} AKA^{-1}$ for any diagonal integer-valued matrix $K$ is just as good of an estimate as $\hat G$.
If we have, e.g., a bound on the norm of $G$, we can remove this ambiguity.
In the remainder of this work we will without loss of generality assume $\smallt = 1$ for brevity, which can be understood as writing $\Omega$ in the units of $\smallt$.

To study the theoretical limits of solving the generator estimation problem, we perform Cramér--Rao analysis for generator estimation, where we take the real and imaginary parts of the additive noise $\eta_{i,j}[\ell]$ to be i.i.d. zero-mean Gaussian noise with variance $\frac{\sigma^2}{2}$. 
For a brief introduction to the Cramér--Rao bound, see~\cref{app:cramer-rao}.
We obtain the following theorem.
\begin{theorem}[Lower bound on the Cramér--Rao bound for generator estimation]\label{thrm:cramer-rao}
    Consider the task of estimating $\Omega$ from data~\cref{eq:intro matrix exponential time-series}, given knowledge of $A$ and $B$, where in the degenerate subspaces of $\Omega$ the rows of $B$ are chosen orthonormal.
    Then, in the asymptotic regime of high signal-to-noise ratio, any estimator $\hat \Omega$ has
    \begin{equation}\label{eq:freq cramer-rao}
        \var[\hat \omega_j - \omega_j] \ge \frac{3\sigma^2}{L(L-1)(2L-1)(A^\dag A)_{j,j}(BB^\dag)_{j,j}}.
    \end{equation}
    Consider the task of estimating $A$ from~\cref{eq:intro matrix exponential time-series} given knowledge of $\Omega$ and $B$ 
    with the same assumption as above.
    Now, the elements of any estimator $\hat A$ satisfy
    \begin{align}\label{eq:gen cramer-rao}
        &\var[\Re(\hat{A}_{j,k}) - \Re(A_{j,k})] \ge \frac{\sigma^2}{2L\left(BB^\dag\right)_{k,k}}
        ,\\
        &\var[\Im(\hat{A}_{j,k}) - \Im(A_{j,k})] \ge \frac{\sigma^2}{2L\left(BB^\dag\right)_{k,k}}.
    \end{align}
    The lower bounds above are achievable by an unbiased estimator up to a constant.
\end{theorem}

\noindent
\begin{proof}
    The assumption about $B$ is without loss of generality for invertible $B$, because of the gauge transformation $A G e^{i\ell\Omega} G^{-1} B$, where $G$ is invertible and block diagonal with blocks corresponding to each degenerate subspace of $\Omega$;
    this assumption simplifies the computation of the Fisher information matrix.
    We have $N^2 L$ independent complex random variables with probability distributions
    \begin{equation}
        p(y_{j,k}[\ell]) = \frac1{\pi\sigma^2} \exp\left[-\frac{\left|y_{j,k}[\ell] - \left(A e^{i\ell\Omega} B\right)_{j,k}\right|^2}{\sigma^2}\right].
    \end{equation}
    In~\cref{app:fisher info details} we compute the Fisher information matrices $F_\Omega$ and $F_A$ corresponding to the estimation of $\omega_k$ and $A$ respectively.
    The diagonal elements of $F_\Omega$ evaluate to
    \begin{equation}\label{eq:diagonal Fisher omega}
        \frac{L(L-1)(2L-1)}{3\sigma^2} \diag(A^\dag A)_{k,k}\diag(BB^\dag)_{k,k},
    \end{equation}
    while the diagonal elements of $F_A$, corresponding to the estimation of $\Re(A_{j,k})$ and $\Im(A_{j,k})$ evaluate to 
    \begin{equation}\label{eq:diagonal Fisher A}
        \frac{2L}{\sigma^2} \diag(BB^\dag)_{k,k}.
    \end{equation}
    The theorem now follows by a direct application of the Cramér--Rao bound, along with
    \begin{equation}
        F^{-1}_{j,j} \ge \frac{1}{F_{j,j}},
    \end{equation}
    which holds due to positive definiteness of the Fisher information matrix.
    Since the probability distribution is periodic in $\Omega$, the Cramér--Rao bound for the estimators $\hat\Omega$ is only valid in the asymptotic regime of a high signal-to-noise ratio~\cite{routtenbergNonBayesianPeriodicCramerRao2013,routtenbergPeriodicCRBNonBayesian2011}.

    We now give estimators achieving these bounds up to a constant.
    For frequency estimation, consider the transformation of the data $y'[\ell] = A^{-1} y[\ell] B^{-1}$. 
    Each frequency can be estimated from the time-series $y'[\ell]_{i,i} = e^{i\ell \omega_i} + (A^{-1} \eta[\ell] B^{-1})_{i,i}$, which is equivalent to estimating a single tone from data with zero-mean Gaussian noise with variance $\sigma^2 (A^{-1}(A^{-1})^\dag)_{i,i}((B^{-1})^\dag B^{-1})_{i,i}$, as can be verified using $\mathbb{E}\eta[\ell]_{i,j} \eta[\ell']_{k,m} = 0$ and $\mathbb{E}\eta[\ell]_{i,j}^* \eta[\ell']_{k,m} = \sigma^2 \delta_{\ell, \ell'} \delta_{i,k} \delta_{j,m}$.
    Maximum likelihood estimation is an asymptotically unbiased estimator achieving variance~\cite{rifeSingleToneParameter1974}
    \begin{equation}
        \var[\hat\omega_i] = \frac{6\sigma^2 (A^{-1}(A^{-1})^\dag)_{i,i}((B^{-1})^\dag B^{-1})_{i,i}}{L(L^2 - 1)},
    \end{equation}
    matching the lower bound~\cref{eq:freq cramer-rao} up to a constant factor.

    To saturate~\cref{eq:gen cramer-rao} up to a constant, consider the unbiased estimator 
    \begin{equation}
        \hat A_0 \coloneqq \frac1L \sum_{\ell\in[L]} y[\ell] B^{-1} e^{-i\Omega \ell}.
    \end{equation}
    The error $E \coloneqq \hat A_0 - A$ is a correlated centred complex Gaussian matrix.
    Using $\mathbb{E}\eta[\ell]_{i,j} \eta[\ell']_{k,m} = 0$ and $\mathbb{E}\eta[\ell]_{i,j}^* \eta[\ell']_{k,m} = \sigma^2 \delta_{\ell, \ell'} \delta_{i,k} \delta_{j,m}$ we can directly evaluate the variance
    \begin{equation}
        \var[\Re(E_{i,j})] = \var[\Im(E_{i,j})] = \frac{\sigma^2}{2L} \left((B^{-1})^\dag B^{-1}\right)_{j,j}.
    \end{equation}
    For $B$ unitary, or at least $B$ such that $BB^\dag$ is diagonal, the estimator $\hat A_0$ matches~\cref{eq:gen cramer-rao} exactly.
\end{proof}

\Cref{thrm:cramer-rao} lower bounds the true Cramér--Rao bound of estimating $\Omega$ and $A$ and the bound is independent of the spectral spacing, working even in the degenerate case.
We can see that the scaling in $L$ of optimal asymptotic precision in estimating $G$ is limited by the precision in estimating its eigenvectors $A$, while its spectrum $\Omega$ can be estimated much more precisely.
This can be understood intuitively: As $L$ increases, more oscillations are visible and so each new sample contains more precise information about the frequencies.
But this is not the case for the distinguishability of eigenvectors of $G$, which oscillates at the frequencies $\Omega$.
Hence the information gain about $A$ from a new sample is nearly constant and the precision in estimating the elements of $\hat A$ decreases with the shot noise scaling $L^{-\frac12}$.

We will now specialise to the case where $A$ and $B$ are unitary matrices, which is the regime in which guarantees for genESPRIT are proven in~\cref{sec:guarantees}.
In this case, we now relate the elementwise variance bounds on $\hat A$ from~\cref{thrm:cramer-rao} to an error on the related propagator and the generator.

\begin{corollary}[Propagator and generator error bounds]\label{cor:gaussian estimator}
    In the case of unitary $A$ and $B$, the estimator $\hat A_0 = \frac1L \sum_{\ell\in[L]} y[\ell] B^{-1} e^{-i\Omega \ell}$ is unbiased and achieves the elementwise variance bound~\cref{eq:gen cramer-rao}.
    The propagator estimate $\hat \Lambda \coloneqq \hat A_0 e^{i\Omega} \hat A_0^\dag$ and the generator estimate $\hat G \coloneqq \hat A_0\Omega \hat A_0^\dag$ satisfy
    \begin{equation}\label{eq:propagator cramer rao}
        \mathbb{E}\left[\|\hat \Lambda - A e^{i\Omega}A^\dag\|_F^2\right] \in \Theta\left(\frac{\sigma^2 N^2}{L}\right)
    \end{equation}
    and
    \begin{equation}
        \mathbb{E}\left[\|\hat G - A \Omega A^\dag\|_F^2\right] \in \Theta\left(\frac{\sigma^2N}{L} \|\Omega\|_F^2\right).
    \end{equation}
\end{corollary}
We prove~\cref{cor:gaussian estimator} in~\cref{app:gaussian estimator}.
Note that, just like~\cref{thrm:cramer-rao},~\cref{cor:gaussian estimator} solves a simpler problem than generator estimation.
Here, we are estimating the generator and the propagator \emph{given} $\Omega$ and $B$.
Although this setting provides useful lower bounds on the Cramér--Rao bound, it comes with some problems.
For instance, if the supplied $\Omega$ is a constant multiple of the identity, the estimators $\hat \Lambda$ and $\hat G$ are clearly suboptimal and we can estimate the generator exactly by setting $\hat G = \Omega$.
The point of~\cref{cor:gaussian estimator} is however to evaluate the performance of an estimation algorithm that is optimal in the sense of~\cref{thrm:cramer-rao} in estimating the propagator and the generator in the case of \emph{generic} $\Omega$.
In fact, because the performance turns out to be independent of the spectrum,~\cref{cor:gaussian estimator} can be understood as a lower bound on the supremum over $\Omega$ of the true Cramér--Rao bound.
We call this the \emph{spectrum-independent Cramér--Rao bound}.
Therefore,~\cref{cor:gaussian estimator} does not exlude the existence of algorithms that find the generator with a better precision in the case of degenerate spectra.
It is important to note, however, that the extent to which this is possible is limited by the spectrum independent bounds on estimating $\Omega$ and $A$ in~\cref{thrm:cramer-rao}.

\section{GenESPRIT algorithm}\label{sec:genESPRIT}
\noindent
A practical algorithm to solve generator estimation has been introduced by the present authors in ref.~\cite{hangleiterRobustlyLearningHamiltonian2024a}.
Here, we coin the algorithm genESPRIT.
\footnote{In ref.~\cite{hangleiterRobustlyLearningHamiltonian2024a} the algorithm has been called tensorESPRIT.
Here we introduce a new name to distinguish it from tensorESPRIT of ref.~\cite{roemerMultidimensionalUnitaryTensorESPRIT2009}, which solves a different problem.}
GenESPRIT is a signal-subspace-based spectral estimation algorithm related to ESPRIT~\cite{paulrajSubspaceRotationApproach1986,royESPRITestimationSignalParameters1989,fannjiangCompressiveSpectralEstimation2016}, suitable for matrix exponential time-series.
We will now give an overview of the algorithm.

For $1<K<L$, consider the block Hankel matrix
\begin{equation}
    \mathcal{H}_K(y) \coloneqq 
    \begin{bmatrix}
        y[0] & y[1] & \dots & y[K-1]\\
        y[1] & y[2] & \dots & y[K]\\
        \vdots && \ddots\\
        y[L-K] & \dots && y[L-1]
    \end{bmatrix}.
\end{equation}
For the noiseless matrix exponential time-series $x[\ell] = Ae^{i\Omega\ell} B$ the Hankel matrix $\mathcal{H}_K(x)$ allows for a block version of the Vandermonde decomposition
\begin{equation}
    \mathcal{H}_K(x) =
    \begin{bmatrix}
        A\\
        A e^{i\Omega}\\
        \vdots\\
        A e^{i\Omega(L-K)}
    \end{bmatrix}
    \begin{bmatrix}
        B & e^{i\Omega} B & \dots & e^{i\Omega(K-1)}B
    \end{bmatrix},
\end{equation}
which shows that $\rank(\mathcal{H}_K(x)) = N$.
Therefore, to denoise the data $y$, we can compute the rank-$N$ truncated \emph{singular value decomposition} (SVD) 
of the Hankel matrix $\mathcal{H}_K(y) \approx U\Sigma V^\dag$ and define 
\begin{equation}
    a[p, q] \coloneqq U[p] \Sigma V[q]^\dag \approx Ae^{i(p+q)\Omega} B,
\end{equation}
where $p$ and $q$ label the $N\times N$ blocks of $U$ and $V^\dag$.
Hence, the matrix
\begin{equation}
    \mathcal{A} \coloneqq \frac1{K(L-K)}\sum_{p \in [L-K], q \in [K]} a[p+1, q] a^+[p, q]
\end{equation}
approximates $A e^{i\Omega} A^{-1}$ and, upon computing its eigendecomposition $\mathcal{A} = X\Lambda X^{-1}$, we obtain the estimates
\begin{equation}
    \hat \Omega \coloneqq \arg \Lambda \quad \text{and} \quad \hat G \coloneqq -i \ln \mathcal{A}
\end{equation}
as a proposed solution to generator estimation, where the principal branch of the logarithm is chosen.
The algorithm is summarised for general $\smallt$ in \cref{alg:genESPRIT}.
Note that if $a[p,q]$ are all invertible, which, as we show in~\cref{sec:guarantees}, will be the case with high probability if $L$ is large enough, we can compute $\mathcal{A}$ via
\begin{equation}\label{eq:better A}
    \mathcal{A} = \frac1{L-K} \sum_{p\in[L-K]} U[p+1]U[p]^{-1},
\end{equation}
reducing the computational effort.

\begin{algorithm}
    \caption{GenESPRIT}
    \label{alg:genESPRIT}
    \begin{algorithmic}
        \REQUIRE $y\in \mathbb{C}^{N\times N \times L}$, $1<K<L$, $\smallt$.
        \STATE Compute the $N$-truncated SVD $\mathcal{H}_K(y) = U \Sigma V^\dag$
        \FOR{$p \in [L-K+1], q\in[K]$}
            \STATE $a[p, q] \gets U[p] \Sigma V[q]^\dag$
        \ENDFOR
        \STATE $\mathcal{A} \gets \frac1{K(L-K)} \sum_{p\in[L-K],q\in[K]} a[p+1, q] a^+[p, q]$ 
        \STATE Eigendecomposition $\mathcal{A} = X \Lambda X^{-1}$
        \STATE $\hat \Omega \gets \frac1{\smallt} \arg(\Lambda)$
        \STATE $\hat G \gets -\frac{i}{\smallt} \ln \mathcal{A}$
        \RETURN $\hat G$, $\hat \Omega$.
    \end{algorithmic}
\end{algorithm}

\subsection{Computational complexity}
\noindent
We now turn to discussing the computational complexity of the algorithm. Computing the truncated SVD of $\mathcal{H}_K(y)$ requires $\mathcal{O}(K(L-K+1)N^3)$ flops~\cite{xuFastAlgorithmsSingular2023}.
To compute $\mathcal{A}$ we need to invert $\mathcal{O}(K(L-K))$ matrices, take $\mathcal{O}(K(L-K))$ matrix products and sums, where the matrix dimensions are $N\times N$, leading to the complexity $\mathcal{O}(K(L-K)N^3)$.
Slightly better scaling in $N$ is possible using optimised methods.
Hence, we find that the time complexity of running genESPRIT scales as $\mathcal{O}(K(L-K) N^3)$.
Its memory complexity has scaling $\mathcal{O}(N^2 L)$, which is of the same order as the size of the input, provided that we use an implicit definition of $\mathcal{H}_K(y)$.

\section{Performance guarantees}\label{sec:guarantees}
\noindent
We now obtain performance guarantees for the genESPRIT algorithm in the case when $A, B$ are unitary.

\begin{theorem}[Performance guarantees for genESPRIT]\label{thrm:genESPRIT guarantees}
    Consider the output $\hat \Omega, \hat G$ of the genESPRIT algorithm with Hankel parameter $1 < K \le \frac{L}2$, given data of the form \cref{eq:intro matrix exponential time-series} with additive noise $\eta[\ell]$, unitary $A$ and $B$,  and $\smallt = 1$.
    Suppose that the noise $\eta[\ell]$ is independent for each $\ell$ and satisfies
    \begin{equation}
    \begin{split}
        \mathbb{E}[\eta[\ell]] = 0&, \ \|\eta[\ell]\|_\infty \le \lambda \ \text{and}\\
        \max(\|\mathbb{E}[\eta^\dag[\ell]\eta[\ell]]\|_\infty&, \|\mathbb{E}[\eta[\ell]\eta^\dag[\ell]]\|_\infty) \le N\sigma^2
    \end{split}
    \end{equation}
    and define the effective noise level $g \coloneqq \frac{f}{\sqrt{P} - f}$, where $P \coloneqq K(L-K+1)$.
    Now, there exist functions 
     \begin{equation}
    \beta \in \mathcal{O}\left(\frac{\sqrt{\sigma^2 N} + \sigma^2N}{\sqrt{L}}\ln \frac{LN}{\delta}\right) 
    \end{equation}
    and $f \in \mathcal{O}\left(\sqrt{\sigma^2 N L\ln \frac{LN}{\delta}}\right)$, such that if $\beta < \sqrt{1 - 2g^2}$ then for any $\delta>0$ with probability $1 - \delta$ the propagator error satisfies
    \begin{equation}\label{eq:generator guarantee}
        \|e^{i\hat G} - e^{iG}\|_F \le \frac{4g\sqrt{2N}}{\sqrt{1 - 2g^2} - \beta} \in \mathcal{O}\left(\frac{\sigma N}{\sqrt{L}}\sqrt{\ln \frac{LN}\delta}\right),
    \end{equation}
    and if furthermore $\frac{2g\sqrt{N}}{\sqrt{1-2g^2}}<1$ then for all $j\in[N]$ there exists an estimate $\hat\omega \in \spec(\hat\Omega)$, such that
    \begin{equation}\label{eq:freq guarantee}
    \begin{split}
         |\hat\omega - \omega_j| &\le \frac{c_0 N g^2}{1 - 2g^2 - \beta\sqrt{1-2g^2}} + \frac{c_1 \beta}{(L-K)\sqrt{1-2g^2}}\\
        &\eqqcolon \xi \ \in \mathcal{O}\left(\left(\frac{\sigma^2 N^2}{L} + \frac{\sigma \sqrt{N}}{L^{\frac32}}\right)\ln \frac{LN}{\delta}\right),
    \end{split}
    \end{equation}
    where $c_0 = 40\pi(e^2 - 1)$ and $c_1=4\pi e^2$ are universal constants, and the matching distance error in the spectrum satisfies
    \begin{equation}\label{eq:freq guarantee md}
        \|\hat \Omega - \Omega\|_{\text{md}, \mathbb{T}} \le N \xi.
    \end{equation}
\end{theorem}

\noindent 
The precise forms of $f$ and $\beta$ are given in~\cref{lemma:noise Hankel,lemma:u v op norm} below.
The conditions $\beta < \sqrt{1 - 2g^2}$ and $\frac{2g\sqrt{N}}{\sqrt{1-2g^2}}<1$ will be satisfied for large enough $L$.

\Cref{thrm:genESPRIT guarantees} shows that the scaling in~\cref{cor:gaussian estimator} \emph{is}, up to logarithmic factors, the actual spectrum-independent Cramér--Rao lower bound for estimating the generator.
In contrast, the scaling in the matching distance error in $\hat \Omega$ does not match~\cref{thrm:cramer-rao}, so it remains open where the actual Cramér--Rao bound lies.

In~\cref{ssec:generator guarantees} we will translate the propagator guarantee~\cref{eq:generator guarantee} to a guarantee on the generator, under a bound on its operator norm known in advance.
However, to do this rigorously, we need to introduce an extra projection step into the algorithm, projecting $\mathcal{A}$ to the unitary group before taking the logarithm.

We leave the proof of guarantees on genESPRIT in the general case of invertible $A$ and $B$ to future research.
We study it numerically in~\cref{sec:numerics} and observe how the performance of genESPRIT becomes worse as the condition numbers of $A$ and $B$ increase.
This is to be expected since the subspaces corresponding to each frequency become harder to distinguish.

In the remainder of the section we prove~\cref{thrm:genESPRIT guarantees}.

\subsection{Bounding the block Hankel matrix of the noise}
\noindent
We first consider the effect of the noise on the block Hankel matrix.
Consider the relation
\begin{equation}
    \mathcal{H}_K(y) = \mathcal{H}_K(x) + \mathcal{H}_K(\eta).
\end{equation}
We would like to use matrix perturbation theory to get a handle on the behaviour of the blocks $U[p]$, $V[q]$ and $\Sigma$ of the SVD of $\mathcal{H}_K(y)$. 
For that, we need to obtain concentration bounds on the operator norm of $\mathcal{H}_K(\eta)$.
Under the noise assumptions in~\cref{thrm:genESPRIT guarantees} we obtain the following lemma, which also defines the precise form of $f$ in~\cref{thrm:genESPRIT guarantees}.
The proof is similar to the proof of Theorem 4 in ref.~\cite{liaoMUSICMultidimensionalSpectral2015}.
\begin{lemma}[Noise Hankel matrix]\label[lemma]{lemma:noise Hankel}
    Given the signal \cref{eq:intro matrix exponential time-series} with additive noise $\eta[\ell]$, which is an independent zero-mean random matrix for each $\ell\in[L]$ with
    \begin{equation}
    \begin{split}
        \max(\|\mathbb{E}[\eta[\ell]^\dag \eta[\ell]]\|_\infty, \|\mathbb{E}[\eta[\ell]\eta[\ell]^\dag]\|_\infty) &\le N\sigma^2,\\
        \|\eta[\ell]\|_\infty &\le \lambda,
    \end{split}
    \end{equation}
    for any $\delta>0$ and integer $1<K\le\frac{L}2$, we have that with probability at least $1 - \frac\delta2$
    \begin{align}\label{eq:def f}
    \begin{split}
        \|\mathcal{H}_K(\eta)\|_\infty &\le \\
        \nonumber
                                       &\hspace{-1.5cm}\ln\left(\frac{2(L+1)N}{\delta}\right) \left(\frac{\lambda}{3} + \sqrt{\frac{\lambda^2}{9} + \frac{2N\sigma^2 (L-K+1)}{\ln \frac{2(L+1)N}{\delta}}}\right)
    \end{split}\\
        &\eqqcolon f \in \mathcal{O}\left(\sqrt{\sigma^2 N L \ln \frac{LN}\delta}\right).
    \end{align}
\end{lemma}
\begin{proof}
We write
\begin{equation}
    \mathcal{H}_K(\eta) = \sum_{\ell \in [L]} \eta[\ell] \odot B^{(\ell)} \eqqcolon \sum_{\ell \in [L]} C^{(\ell)},
\end{equation}
where $B^{(\ell)} \in \mathbb{C}^{N(L-K+1) \times NK}$ is the matrix, which has the $N\times N$ identity matrices everywhere on the $\ell$-th secondary $N\times N$-block diagonal and zeroes everywhere else, and the $\odot$ operation premultiplies each $N\times N$ block of $B^{(\ell)}$ by $\eta[\ell]$ via matrix multiplication.
With this decomposition, the matrix Bernstein inequality (Theorem 1.6.2 of ref.~\cite{troppIntroductionMatrixConcentration2015}) ensures that with probability at least $1-\frac\delta2$
\begin{equation}
    \|\mathcal{H}_K(\eta)\|_\infty \le \ln \frac{2(L+1)N}{\delta} \left(\frac{\lambda}{3} + \sqrt{\frac{\lambda^2}{9} + \frac{2\nu}{\ln \frac{2(L+1)N}\delta}} \right),
\end{equation}
where 
\begin{equation}
    \nu \coloneqq \max\{\|\sum_\ell \mathbb{E}[(C^{(\ell)})^\dag C^{(\ell)}]\|_\infty, 
    \|\sum_\ell \mathbb{E}[C^{(\ell)} (C^{(\ell)})^\dag]\|_\infty \}.
\end{equation}
Using the assumptions on the 
noise, we find that
\begin{align}
    \|\sum_\ell \mathbb{E}[(C^{(\ell)})^\dag C^{(\ell)}]\|_\infty &\le N\sigma^2 (L-K+1),\\
    \|\sum_\ell \mathbb{E}[C^{(\ell)} (C^{(\ell)})^\dag] \|_\infty &\le N\sigma^2 K,
\end{align}
so that $\nu \le N \sigma^2 (L-K+1)$, obtaining the lemma.
\end{proof}

\subsection{Properties of blocks of singular vector matrices}
\noindent
Having control over $\mathcal{H}_K(\eta)$ we are now in the position to use perturbation theory to understand the effect of noise on the $N \times N$ blocks of $U$ and $V$, the rank-$N$ truncated singular matrices of $\mathcal{H}_K(y)$.
The main tool we need is Wedin's theorem.

\begin{theorem}[Wedin's theorem \cite{wedinPerturbationBoundsConnection1972}]\label{thrm:wedin}
    Suppose matrices $M, \tilde M \in \mathbb{C}^{n\times m}$, where $\rank(M)=r$, with rank-$r$ truncated singular value decompositions $M = U\Sigma V^\dag$ and $\tilde M = \tilde U \tilde \Sigma \tilde V^\dag$ and let $\Delta \coloneqq \tilde M - M$.
    Then for all $j \in [r]$ the singular values satisfy
    \begin{align}
        \sigma_{r-1}(M) - \|\Delta\|_\infty \le \sigma_j(\tilde M) \le \sigma_0(M) + \|\Delta\|_\infty
    \end{align}
    and, given $\|\Delta\|_\infty < \sigma_{r-1}(M)$, the singular values of the matrix $U^\dag \tilde U$ lie in the interval
    \begin{equation}
        \left[\sqrt{1 - \frac{2\|\Delta\|_\infty^2}{(\sigma_{r-1}(M) - \|\Delta\|_\infty)^2}} \ , \ 1\right],
    \end{equation}
    while the singular values of the matrix $U^\dag_\perp \tilde U$, where $U_\perp$ is chosen such that $\begin{pmatrix}U & U_\perp\end{pmatrix}$ is unitary, are all upper bounded by $\frac{\sqrt{2}\|\Delta\|_\infty}{\sigma_{r-1}(M) - \|\Delta\|_\infty}$.
\end{theorem}

\noindent In our case, we identify $M$ with $\mathcal{H}_K(x) = U_0 \Sigma_0 V_0^\dag$ and $\tilde M$ with $\mathcal{H}_K(y)$.
Since $A$ and $B$ are assumed unitary, we can choose
\begin{align}
    U_0 = 
    \frac1{\sqrt{L-K+1}}&
    \begin{bmatrix} 
        A\\
        Ae^{i\Omega}\\
        \vdots\\
        Ae^{i\Omega(L-K)}
    \end{bmatrix},\\
    V_0 = \frac1{\sqrt{K}}
    \begin{bmatrix}
        B^\dag\\
        B^\dag e^{-i\Omega}\\
        \vdots\\
        B^\dag e^{-i\Omega (K-1)}
    \end{bmatrix}\ &\text{and} \ \Sigma_0 = \sqrt{P} \ \id
\end{align}
as the SVD of $\mathcal{H}_K(x)$, where $P \coloneqq K(L-K+1)$.
To use Wedin's theorem to obtain information about the $N\times N$ blocks $U[p]$ and $V[q]$ of $U$ and $V$, respectively, consider the matrix 
\begin{equation}\label{eq:omega U definition}
    \overline \omega_U \coloneqq U_0^\dag U = \frac1{\sqrt{L-K+1}} \sum_{p\in[L-K+1]} e^{-i\Omega p} A^\dag U[p],
\end{equation}
and $\overline \omega_V \coloneqq V_0^\dag V$, which are close to a unitary due to \cref{thrm:wedin}.
In the case of no noise, $\overline \omega_U$ would be exactly unitary and it would encode the gauge freedom in the SVD decomposition; we would have $U[p] = (L-K+1)^{-\frac12} A e^{i\Omega p} \overline \omega_U$.
In the noisy case, we write $U[p]$ and $V[q]$ as perturbations of this relation 
\begin{equation}\label{eq:u v ansatz}
    \begin{split}
        U[p] &= \frac{1}{\sqrt{L-K+1}} A e^{i\Omega p} (\overline \omega_U + \delta_U[p]),\\
        V[q] &= \frac{1}{\sqrt{K}} B^\dag e^{-i\Omega q} (\overline \omega_V + \delta_V[q]),
    \end{split}
\end{equation}
and in the following lemma use Wedin's theorem to bound the deviations $\delta_U[p]$ and $\delta_V[q]$.
\begin{lemma}[Perturbation bounds for singular-vector blocks]\label[lemma]{lemma:u v blocks}
    Let $\delta_U$ and $\delta_V$ be as in eq.~\cref{eq:u v ansatz} and $\overline \omega_U \coloneqq U_0^\dag U, \ \overline \omega_V \coloneqq V_0^\dag V$.
    Now, for any $\delta>0$ with probability at least $1 - \frac\delta2$
    \begin{equation}
        1 \ge \sigma_j(\overline\omega_{U,V}) \ge \sqrt{1- 2 g^2},
    \end{equation}
    where $g \coloneqq f(\sqrt{P} - f)^{-1}$, $f$ is defined in~\cref{eq:def f} and $P \coloneqq K(L-K+1)$.
    Furthermore $\sum_p \delta_U[p] = \sum_q \delta_V[q] = 0$ and
    \begin{equation}
    \begin{split}
        \sum_p \|\delta_U[p]\|_F^2 &\le 2N(L-K+1) g^2 \label{eq:delta frob norm condition}, \\
        \sum_q \|\delta_V[q]\|_F^2 &\le 2NKg^2.
    \end{split}
    \end{equation}
\end{lemma}
\begin{proof}
    Plugging \cref{eq:u v ansatz} into \cref{eq:omega U definition}, using the unitarity of $A$ we get that $\sum_{p\in[L-K+1]} \delta_U[p] = 0$.
    Using this fact and inserting \cref{eq:u v ansatz} into the isometry property $U^\dag U = \id_N$, we obtain
    \begin{align}\label{eq:sum of deltas}
        \sum_{p\in[L-K+1]} \delta_U[p]^\dag \delta_U[p] = (L-K+1) \left(\id - \overline \omega_U^\dag \overline \omega_U\right).
    \end{align}
    Taking the trace, \cref{thrm:wedin} gives us 
    \begin{equation}
        \sum_{p\in[L-K+1]} \|\delta_U[p]\|_F^2 \le \frac{2N\|\mathcal{H}_K[\eta]\|_\infty^2}{K\left(1 - \frac{\|\mathcal{H}_K(\eta)\|_\infty}{\sqrt{P}}\right)^2}.
    \end{equation}
    From \cref{lemma:noise Hankel} we hence obtain that with probability $1 - \frac\delta2$ \cref{eq:delta frob norm condition} holds.
    Analogously, we have
    \begin{equation}
        \overline \omega_V \coloneqq V_0^\dag V
    \end{equation}
    and we can write for each $q\in[K]$ the ansatz~\cref{eq:u v ansatz}.
    Following the same steps as before, we prove the rest of \cref{lemma:u v blocks}.
\end{proof}

Next we will derive concentration inequalities on $\max_{p\in[L-K+1]}\|\delta_U[p]\|_\infty$ and $\max_{q\in[K]}\|\delta_V[q]\|_\infty$.
The idea is to note that $U = \mathcal{H}_K(y) V \Sigma^{-1}$ and write $\mathcal{H}_K(y) = \sqrt{P} U_0 V_0^\dag + \mathcal{H}_K(\eta)$, such that
\begin{equation}
    U[p] = \sqrt{P} U_0[p] \overline \omega_V \Sigma^{-1} + h_K(\eta)[p] V \Sigma^{-1},
\end{equation}
where $h_K(\eta)[p]$ is the $p$-th block row of $\mathcal{H}_K(\eta)$.
The first term is well-conditioned due to~\cref{thrm:wedin}, while the second term concentrates due to the matrix Bernstein inequality (Theorem 1.6.2 in ref.~\cite{troppIntroductionMatrixConcentration2015}).
Now, comparing with~\cref{eq:u v ansatz} we can bound the operator norm of $\delta_U[p]$ with high probability.
Similar logic can be applied to bound the operator norm of $\delta_V[q]$.
Using the union bound, we can bound the maxima and get the following lemma, which is proven in~\cref{app:u v op norm}.
\begin{lemma}[Uniform operator-norm bounds for block perturbations]\label{lemma:u v op norm}
    The matrices $\delta_U[p]$ and $\delta_V[q]$ from~\cref{lemma:u v blocks} satisfy for any $\delta>0$ with probability at least $1 - \delta$, on top of the bounds in~\cref{lemma:u v blocks},
    \begin{equation}
    \begin{split}
        \max_{p\in[L-K+1]} \|\delta_U[p]\|_\infty &\le \beta,\\
        \max_{q\in[K]} \|\delta_V[q]\|_\infty &\le \beta,
    \end{split}
    \end{equation}
    where
    \begin{equation}
        \beta \coloneqq g\left(1 + g \sqrt{2(L-K+1)}\right) + \frac\gamma{\sqrt{K}\left(1 - \frac{f}{\sqrt{P}}\right)}
    \end{equation}
    with
    \begin{equation}
        \gamma \coloneqq \ln\frac{4NL}{\delta} \left(\frac{4\lambda}{3\sqrt{K}} + \sqrt{\frac{16 \lambda^2}{9 K} + \frac{8N\sigma^2}{\ln \frac{4NL}{\delta}}}\right),
    \end{equation}
    such that $\beta \in \mathcal{O}\left(\frac{\sqrt{\sigma^2 N} + \sigma^2 N}{\sqrt{L}} \ln \frac{LN}{\delta}\right)$.
\end{lemma}

We can see that blocks of $U$ and $V$ will be invertible with probability $1 - \delta$ whenever $\beta < \sqrt{1 - 2g^2}$, which can be guaranteed for large enough $L$.

\subsection{Guarantees for the propagator}
\noindent
We now have all the technical tools necessary to compute the guarantee on the propagator.
In the event when the blocks $U[p]$, $V[q]$ and $\Sigma$ are invertible, we have $a[p+1, q] a^+[p,q] = U[p+1]U[p]^{-1}$ and we can write
\begin{align}
    \mathcal{A} &= \frac1{L-K}\sum_{p\in[L-K]} U[p+1] U[p]^{-1}\\
    \begin{split}
    &= \frac1{L-K}\sum_{p\in[L-K]} A e^{i\Omega(p+1)} (\overline{\omega}_U + \delta_U[p+1]) \\
    &\hspace{1cm}\times(\overline \omega_U + \delta_U[p])^{-1} e^{-i\Omega p} A^\dag.
    \nonumber
    \end{split}
\end{align}
Using the identity
\begin{equation}
\begin{split}
    (\overline \omega_U + \delta_U[p+1])&(\overline \omega_U + \delta_U[p])^{-1} - \id =\\
                          &(\delta_U[p+1] - \delta_U[p])(\overline \omega_U + \delta_U[p])^{-1},
\end{split}
\end{equation}
which can be verified by post-multiplying with $(\overline \omega_U + \delta_U[p])$, we find that
\begin{equation}
    \mathcal{A} = Ae^{i\Omega} A^\dag + \Delta \mathcal{A},
\end{equation}
where
\begin{equation}
\begin{split}
    \Delta\mathcal{A} = \frac1{L-K} \sum_{p\in[L-K]} A &e^{i\Omega (p+1)} (\delta_U[p+1] - \delta_U[p])\times\\
    &\times(\overline \omega_U + \delta_U[p])^{-1} e^{-i\Omega p} A^\dag.
\end{split}
\end{equation}
This term can be bounded using subaditivity of the operator norm,~\cref{eq:delta frob norm condition} and the Cauchy--Schwarz inequality, obtaining
\begin{align}
    \|\Delta\mathcal{A}\|_F &\le \frac2{\sqrt{L-K}(\sqrt{1 - 2g^2} - \beta)} \sqrt{\sum_{p\in[L-K]} \|\delta_U[p]\|_F^2}\nonumber \\
    &\le \frac{4g\sqrt{2N}}{\sqrt{1 - 2g^2} - \beta} \in \mathcal{O}\left(\sqrt{\frac{\sigma^2N^2}{L}\ln \frac{N}\delta}\right)
\end{align}
which proves~\cref{eq:generator guarantee}.

\subsection{Guarantees for the generator}\label{ssec:generator guarantees}

In some applications the meaningful figure of merit is the error in the generator itself, rather than the error in the propagator~\cref{eq:generator guarantee}.
We would therefore like to translate the guarantee~\cref{eq:generator guarantee} to a guarantee on the logarithm of $\mathcal{A}$.
For this to be possible, we need to assume that $\|\Omega\|_\infty \le \frac\pi2$, such that $G$ lies in the principal branch of the logarithm and such that the (scalar) logarithm has a tractable Lipschitz constant $\frac\pi2$ on the domain of the spectrum of $e^{iG}$.

We would now like to translate Lipschitz continuity of the scalar logarithm to the matrix case.
To this end, we utilise the main result of ref.~\cite{kittanehLipschitzFunctionsNormal1985}, 
which, however, 
requires $\mathcal{A}$ to be a normal matrix.
To guarantee this, we will deviate from~\cref{alg:genESPRIT} slightly and include the extra step of computing the projection $\mathcal{A}'$ of $\mathcal{A}$ to the unitary group and let $\hat G = -i\ln \mathcal{A}'$, where we use the principal branch of the logarithm.
Labelling the error~\cref{eq:generator guarantee} by $\epsilon$, we have that $\|\mathcal{A}' - A e^{i\Omega} A^\dag \|_F \le 2\epsilon$ with probability $1-\delta$.
From the assumption $\|\Omega\|_\infty \le \frac\pi2$ we show in~\cref{app:from exp to generator} that with probability $1 - \delta$ the estimate $\hat G$ satisfies $\|\hat G\|_\infty \le \frac\pi2 + \theta$, where
\begin{equation}
    \theta = \frac{2\epsilon}{\sqrt{1 - \left(\frac1{\sqrt{2}} + \epsilon\right)^2}},
\end{equation}
given that $\frac1{\sqrt{2}} + \epsilon < 1$.
If $\theta < \frac\pi2$, for unit complex numbers with argument in $[-\frac\pi2-\theta, \frac\pi2+\theta]$ the scalar logarithm has Lipschitz constant $\frac{\pi+2\theta}{2\cos\theta}$.
Hence, using the results of 
ref.~\cite{kittanehLipschitzFunctionsNormal1985} we obtain the guarantee on the generator
\begin{equation}\label{eq:generator error}
    \|\hat G - A \Omega A^\dag\|_F \le \frac{2(\pi+2\theta)}{2\cos\theta} \epsilon.
\end{equation}

\subsection{Guarantees for the spectrum}

Equipped with a bound on $\|\Delta\mathcal{A}\|_F$ from the previous section, we already have a handle on the eigenvalue error through the Bauer--Fike theorem and $\|\cdot\|_\infty \le \|\cdot \|_F$.
However, the scaling in $L$ so obtained gives (up to logarithmic factors) $L^{-\frac12}$, which is a whole factor of $L$ worse than what the lower bound~\cref{thrm:cramer-rao} provides.
Can we do better?
In this section we show that due to the specific structure of $\Delta\mathcal{A}$ we actually can, obtaining an improved scaling $L^{-1}$ (up to logarithmic factors).

First, notice the identity
\begin{equation}
\begin{split}
    (\overline \omega &+ \delta_U[p+1])(\overline \omega_U + \delta_U[p])^{-1} - \id = \\
                      &(\delta_U[p+1] - \delta_U[p]) \overline\omega_U^{-1} - (\delta_U[p+1] - \delta_U[p]) \overline\omega_U^{-1} \times \\ 
                      &\hspace{3cm}\times\delta_U[p](\overline\omega_U + \delta_U[p])^{-1},
\end{split}
\end{equation}
which can be verified by post-multiplying with $(\overline\omega_U + \delta_U[p])$.
Now we can decompose $\Delta \mathcal{A}$ as $\Delta \mathcal{A} = \mathcal{A}_{[]} + \mathcal{A}_0 - \mathcal{A}_U$, where
\begin{align}
    \mathcal{A}_{[]} &\coloneqq \frac1{L-K} \sum_{p=1}^{L-K-1} A e^{i\Omega p} [\delta_U[p]\overline\omega_U^{-1}, e^{i\Omega}] e^{-i\Omega p} A^\dag ,\\
    \begin{split}
        \mathcal{A}_0 &\coloneqq \frac1{L-K} Ae^{i\Omega(L-K)} \delta_U[L-K] \overline \omega_U^{-1} e^{-i\Omega (L-K-1)} A^\dag - \\
                      &\hspace{2cm} \frac1{L-K} A e^{i\Omega} \delta_U[0] \overline\omega_U^{-1} A^\dag.
    \end{split}\\
    \begin{split}
        \mathcal{A}_U &\coloneqq \frac1{L-K} \sum_{p\in[L-K]} A e^{i\Omega(p+1)} (\delta_U[p+1] - \delta_U[p]) \times\\
        &\hspace{1cm}\times\overline\omega_U^{-1} \delta_U[p](\overline\omega_U + \delta_U[p])^{-1} e^{-i\Omega p} 
        A^\dag.
    \end{split}
\end{align}
Using~\cref{lemma:u v blocks,lemma:u v op norm}, 
we can bound
\begin{equation}
    \|\mathcal{A}_0\|_\infty \le \frac{2\beta}{(L-K)\sqrt{1-2g^2}}
\end{equation}
giving us an $L^{-\frac32}$ scaling.
Submultiplicativity and subadditivity of the operator norm, along with $\|\cdot\|_\infty \le \|\cdot \|_F$, gives
\begin{equation}\label{eq:AU working}
    \begin{split}
        \|\mathcal{A}_U\|_\infty &\le \frac1{L-K} \sum_{p\in[L-K]} (\|\delta_U[p+1]\|_F + \|\delta_U[p]\|_F)\times\\
        &\quad\times \|\delta_U[p]\|_F \|\overline \omega_U^{-1}\|_\infty \|(\overline \omega_U + \delta_U[p])^{-1}\|_\infty.
    \end{split}
\end{equation}
We can bound $\|\overline\omega_U^{-1}\|_\infty \le \frac1{\sqrt{1-2g^2}}$ using~\cref{lemma:u v blocks} and $\|(\overline \omega_U + \delta_U[p])^{-1}\|_\infty \le (\sqrt{1 - 2g^2} - \beta)^{-1}$ using~\cref{lemma:u v op norm}, which both hold with probability at least $1-\delta$.
The remaining factors in eq.~\cref{eq:AU working} are bounded by~\cref{lemma:u v blocks} and the Cauchy--Schwarz inequality.
We get that with probability at least $1-\delta$
\begin{equation}
    \|\mathcal{A}_U\|_\infty \le \frac{8Ng^2}{1 - 2g^2 - \beta\sqrt{1-2g^2}},
\end{equation}
which scales as $\mathcal{O}\left(\frac{\sigma^2N^2}{L}\ln \frac{LN}{\delta}\right)$, matching the target scaling in $L$.

We will get the improvement in the $\mathcal{A}_{[]}$ term by exploiting the commutator structure.
Writing it in 
the $A$ basis, we get
\begin{align}
\nonumber
    \left(A^\dag \mathcal{A}_{[]} A\right)_{j,k} &= \frac{e^{i \omega_k} - e^{i \omega_j}}{L-K} \sum_{p=1}^{L-K-1} 
    e^{i (\omega_j - \omega_k) p} \left(\delta_U[p] \overline \omega_U^{-1}\right)_{j,k}\\
                                        &\eqqcolon (e^{i \omega_j} - e^{i \omega_k}) H_{j,k} = [e^{i\Omega}, H]_{j,k},
\end{align}
where we can bound
\begin{align}
    \|H\|_\infty &\le \frac1{(L-K)\sqrt{1-2g^2}} \sum_{p\in[L-K]} \|\delta_U[p]\|_\infty\\
                 &\le \frac1{\sqrt{(L-K)(1-2g^2)}} \sqrt{\sum_{p\in[L-K]} \|\delta_U[p]\|_F^2}
                 \nonumber\\
                 &\le \frac{2g \sqrt{N}}{\sqrt{1 - 2g^2}}
                 \nonumber
\end{align}
using the Cauchy--Schwarz inequality in the second line.
While the norm $\|H\|_\infty$ shows the $L^{-\frac12}$ scaling as before, we can get an improvement using the following lemma.

\begin{lemma}[Second-order eigenvalue bounds for commutator perturbations]\label[lemma]{lemma:commutator eigenvalue perturbation}
    Consider a matrix $M$ with $\|M\|_\infty = 1$, which is diagonalisable by $Q$ with condition number $\kappa(Q)$, and 
    a perturbation $\widetilde M = M + [M, X] + Y$, where the matrices $X, Y$ satisfy $\|X\|_\infty \le \epsilon < 1$ and $\|Y\|_\infty \le \tilde \epsilon$.
    Now, for any eigenvalue $\tilde \lambda \in \spec(\widetilde{M})$ there exists an eigenvalue $\lambda \in \spec(M)$, such that
    \begin{equation}\label{eq:commutator perturbation spec}
        |\tilde{\lambda} - \lambda| \le \kappa(Q)\left((3e^2 - 5) \epsilon^2 + e^{2} \tilde \epsilon\right).
    \end{equation}
    Furthermore, if $M$ is a normal $N \times N$ matrix, then
    \begin{equation}\label{eq:commutator perturbation spec md}
        \|\spec(\widetilde M) - \spec(M)\|_\text{md} \le N\left((3e^2 - 5) \epsilon^2 + e^2 \tilde\epsilon\right).
    \end{equation}
\end{lemma}
\begin{proof}
    Taylor expanding the matrix exponential, we get
    \begin{align}
        \begin{split}
        e^X \widetilde M e^{-X} &= M + e^X Y e^{-X} + \sum_{k=2}^\infty \frac{X^k M}{k!} + \\
                     &\hspace{-1cm}+ \sum_{k'=2}^\infty (-1)^{k'} \frac{M X^{k'}}{k'!} + \sum_{k, k' = 1}^\infty (-1)^{k'} \frac{X^k M X^{k'}}{k!k'!}+\\
                     &\hspace{-1cm}+ \sum_{k=1}^\infty \left[\frac{X^kM}{k!},X\right] + \sum_{k'=1}^\infty (-1)^{k'}\left[\frac{M X^{k'}}{k!}, X\right] + \\
                     &\hspace{-1cm}+ \sum_{k,k'=1}^\infty \left[\frac{X^k M X^{k'}}{k!k'!}, X\right],
    \end{split}
    \end{align}
    where notably the terms linear in $X$ cancel.
    Now we bound for any matrix $A$
    \begin{align}
        \left\|\sum_{k=1}^\infty \frac{X^k A}{k!}\right\|_\infty \le &\sum_{k=1}^\infty \frac{\epsilon^k}{k!} \|A\|_\infty = (e^\epsilon - 1)\|A\|_\infty ,\\
        \left\|\sum_{k'=1}^\infty (-1)^{k'} \frac{A X^{k'}}{k'!} \right\|_\infty \le &\sum_{k'=1}^\infty \frac{\epsilon^{k'}}{k'!} \|A\|_\infty = (e^{\epsilon} - 1)\|A\|_\infty ,\\
        \begin{split}
            \left\|\sum_{k, k' = 1}^\infty (-1)^{k'} \frac{X^k A X^{k'}}{k!k'!}\right\|_\infty &\le \sum_{k=1}^\infty \frac{\epsilon^k}{k!} \sum_{k'=1}^\infty \frac{\epsilon^{k'}}{k'!} \|A\|_\infty \\
            &= (e^\epsilon - 1)^2\|A\|_\infty.
        \end{split}
    \end{align} 
    Hence, defining $R \coloneqq e^X \widetilde M e^{-X} - M$, we have
    \begin{equation}
        \|R\|_\infty \le 2(e^\epsilon - 1 - \epsilon) + (e^\epsilon - 1)^2 + 4(e^\epsilon - 1)\epsilon + 2(e^\epsilon - 1)^2\epsilon + e^{2\epsilon}\tilde \epsilon.
    \end{equation}
    Consider the function $h(x) \coloneqq e^x - 1 - x - (e-2)x^2$.
    As $h(0) = h(1) = h'(0) = 0$, while $h^{\prime\prime}(0) < 0$ and $h^{\prime \prime \prime}(x) > 0 \ \forall x\in\mathbb{R}$, we know that $h(x) \le 0$ on $[0, 1]$.
    Similarly, the function $\tilde{h}(x) \coloneqq (e^x - 1)^2 - (e-1)^2 x^2$ has $\tilde{h}(0) = \tilde{h}(1) = \tilde{h}'(0) = 0$, while $\tilde{h}^{\prime\prime}(0) < 0$ and $\tilde{h}^{\prime\prime\prime}(x) > 0 \ \forall x \in [0, \infty)$, so that $\tilde{h}(x) \le 0$ on $[0, 1]$.
    Hence, on $\epsilon \in [0, 1]$ we have $e^\epsilon - 1 - \epsilon \le (e - 2) \epsilon^2$ and $(e^\epsilon - 1)^2 \le (e - 1)^2 \epsilon^2$.
    Therefore, we have that 
    \begin{equation}
        \|R\|_\infty \le (e^2 + 1)\epsilon^2 + e^{2\epsilon}\tilde{\epsilon} + 2(e^2 - 3)\epsilon^3 \le (3e^2 - 5) \epsilon^2 + e^2 \tilde \epsilon.
    \end{equation}
    Since $e^X M e^{-X}$ is a similarity transform, Bauer--Fike theorem gives~\cref{eq:commutator perturbation spec} and Corollary 1.2 from ref.~\cite{sunVariationSpectrumNormal1996} gives~\cref{eq:commutator perturbation spec md}, completing the proof. 
\end{proof}

In the case that
\begin{equation}
    \frac{2g\sqrt{N}}{\sqrt{1-2g^2}} < 1
\end{equation}
we can apply~\cref{lemma:commutator eigenvalue perturbation} to $A^\dag \mathcal{A} A = e^{i\Omega} + [e^{i\Omega}, H] + A^\dag \mathcal{A}_0 A + A^\dag \mathcal{A}_U A$, obtaining that for any $j\in[N]$ there is, with probability at least $1-\delta$, an eigenvalue $\lambda$ of $\mathcal{A}$ such that
\begin{equation}
\begin{split}
    |\lambda - e^{i\omega_j}| \le &\frac{4(3e^2 - 5) N g^2}{1 - 2g^2} + \frac{8e^2 N g^2}{1 - 2 g^2 - \beta \sqrt{1 - 2g^2}}+\\
    &\hspace{1.5cm} + \frac{2e^2 \beta}{(L-K)\sqrt{1-2g^2}}\\
    &\hspace{-1.5cm}\le \frac{20(e^2 - 1)N g^2}{1 - 2g^2 - \beta\sqrt{1 - 2g^2}} + \frac{2e^2 \beta}{(L-K)\sqrt{1 - 2g^2}}\eqqcolon \frac{\xi}{2\pi}
\end{split}
\end{equation}
The frequency estimate $\hat \omega \coloneqq \arg \lambda$ in the worst case doubles the error to $|e^{i\hat \omega} - e^{i\omega_j}| \le \pi^{-1} \xi$.
We translate the eigenvalue error to the frequency error using eq. (3.1) in ref.~\cite{liSuperResolutionLimitESPRIT2020}, which gives
\begin{equation}
    |\hat \omega - \omega_j| \le \pi |e^{i\hat \omega} - e^{i\omega_j}| \le \xi,
\end{equation}
proving~\cref{eq:freq guarantee}.
By the same procedure we can prove the matching distance bound~\cref{eq:freq guarantee md} by using~\cref{eq:commutator perturbation spec md}, which gains an extra factor of $N$.

\section{Numerical experiments}\label{sec:numerics}
\noindent
With the theoretical guarantees in place, in this section we assess the performance of genESPRIT numerically. 
We use the following noise model for the elements $y_{j,k}[\ell]$:
write $y_{j,k}[\ell] = 2n^{-1}b_{j,k}[\ell] - 1 - i$, where the real (imaginary) part of $b_{j,k}[\ell]$ is drawn from a binomial distribution $B(n, p_{j,k}[\ell])$, where $n$ is the number of shots, which we use to control the error, and the mean $p_{j,k}[\ell]$ is chosen such that $\mathbb{E}[y_{j,k}[\ell]] = (A e^{i\ell\Omega} B)_{j,k}$. 
Hence, we have that $\lambda = 2N$ and $\sigma = \sqrt{2n^{-1}}$. 
To study a representative part of the problem space of generator estimation, we draw $A$ and $B$ from the Haar measure over the $N\times N$ unitary group and we draw $\Omega$ from the uniform distribution over $(-\pi, \pi]$.

\begin{figure}
    \centering
    \includegraphics[width=\columnwidth]{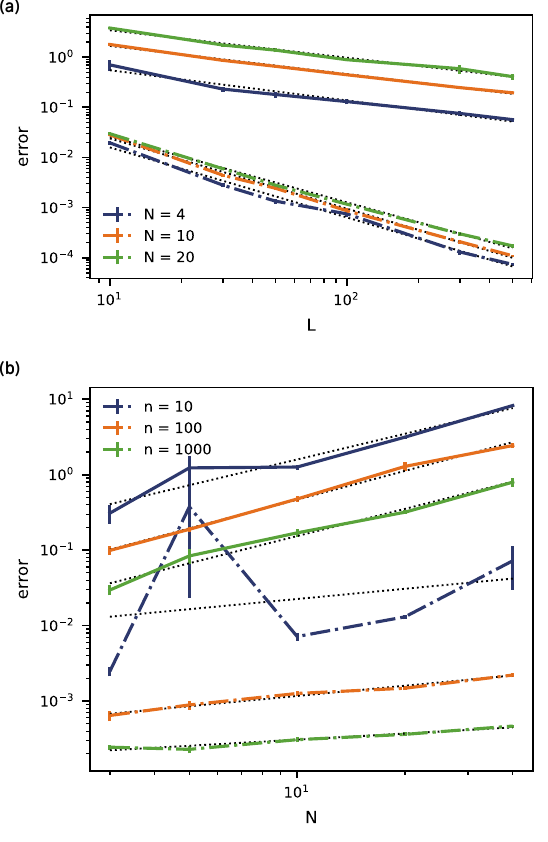}
    \caption{\textbf{Performance of genESPRIT under shot noise.} The generator Frobenius norm error (line) and the matching distance error in the frequencies (dash-dot) are plotted as a function of $L$ with $n=100$ and $N=4, 10, 20$ \textbf{(a)} and $N$ with $L=80$ and $n=10, 100, 1000$ \textbf{(b)}. Each data point is an average of $10$ independent draws of $\Omega$, $A$ and $B$. Lines of best fit are shown and gradients are estimated in \cref{table:scaling exponents}.}
    \label{fig:error scaling}
\end{figure}

We evaluate the numerical performance of genESPRIT by calculating the matching distance error on the eigenvalues and the Frobenius norm in the generator.
We compare the error in the generator with our guarantee~\cref{eq:generator error}, even though the guarantee technically requires a modification of the genESPRIT algorithm to include an additional projection step (see~\cref{ssec:generator guarantees}), which we do not do in the numerical implementation, to see whether this is necessary in practice.
The numerical results are shown in \cref{fig:error scaling} and the empirical scaling in $L$ and $N$, obtained from the line of best fit on a log-log plot, is shown in~\cref{table:scaling exponents}.

We can see that the scaling in $L$ of the error in $\hat G$ matches~\cref{eq:generator error}, while the scaling in $L$ of the error in $\hat\Omega$ is numerically consistently better than the guarantee~\cref{eq:freq guarantee}, even though not quite achieving the lower bound from~\cref{thrm:cramer-rao}.
Considering the scaling in $N$, the error in $\hat G$ is consistent with~\cref{eq:generator error}, as $\lambda$ scales as $\mathcal{O}(N)$ in this case and we are not in the large $L$ limit ($L=80$ here), such that the scaling in $\lambda$ dominates the scaling of $f$ (see~\cref{lemma:noise Hankel}).
The scaling with $N$ of the error in $\hat\Omega$ at $n=100$ agrees with the low noise limit of~\cref{eq:freq guarantee}, but we do not numerically observe the extra factor of $N$ in the matching distance guarantee~\cref{eq:freq guarantee md}. 
At $n=1000$ the empirical scaling is considerably better than both bounds.
We never observe the scaling in $N$ at the second order in the noise in the guarantees~\cref{eq:freq guarantee,eq:freq guarantee md}.
There is a significant outlier in the error in $\hat \Omega$ at $n=10, N=5$, panel (b) of~\cref{fig:error scaling}.

\begin{figure}
    \centering
    \includegraphics[width=\columnwidth]{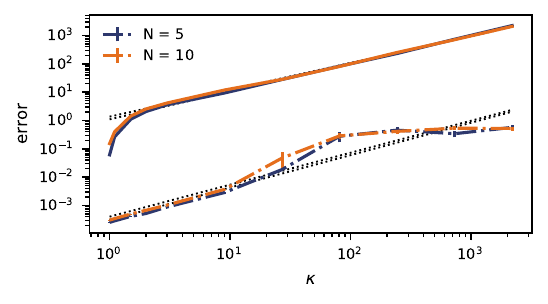}
    \caption{\textbf{Performance of genESPRIT with varying condition number of $A$ and $B$.} The generator Frobenius norm error (line) and the matching distance error in the frequencies (dash-dot) are plotted as a function of the condition number $\kappa = \kappa(A) = \kappa(B)$ with $n=1000$, $L = 80$ and $N=5, 10$. Each data point is an average of $50$ independent draws of $\Omega$, $A$ and $B$. Lines of best fit are shown and gradients are estimated in \cref{table:scaling exponents}.
    For the error in the generator, the data are fitted from $\kappa\ge 2$ where the behaviour is close to linear.}
    \label{fig:scaling kappa}
\end{figure}

In~\cref{fig:scaling kappa} we show the deterioration of the performance of genESPRIT as the condition numbers of $A$ and $B$ increase.
For each $\kappa$ we set the largest (smallest) singular value of $A$ and $B$ to $1$ ($\kappa^{-1}$) and draw the remaining singular values from the uniform distribution over $[\kappa^{-1}, 1]$.
We then draw the singular vectors of $A$ and $B$ from the Haar measure over the unitary group.
The plot is averaged over $50$ instances with shot noise at $n=1000$.
We can see that the behaviour is not linear, especially for the generator error.
We fit a straight line to the approximately linear part of the data and show the resulting scaling in~\cref{table:scaling exponents}.

\begin{figure}
    \centering
    \includegraphics[width=\columnwidth]{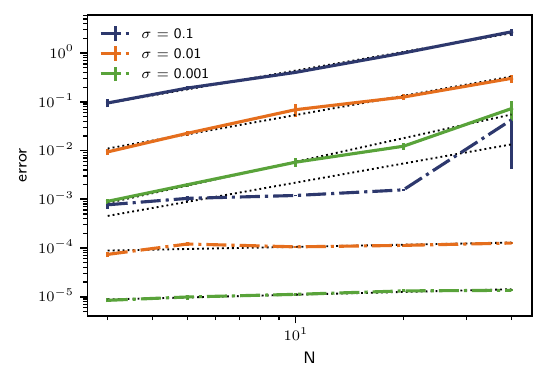}
    \caption{\textbf{Performance of genESPRIT under Gaussian noise.} The generator Frobenius norm error (line) and the matching distance error in the frequencies (dash-dot) are plotted as a function of $N$ with $L=80$ and $\sigma=0.1, 0.01, 0.001$.
    Each data point is an average of $10$ independent draws of $\Omega$, $A$ and $B$.
    Lines of best fit are shown and gradients are estimated in \cref{table:scaling exponents gauss}.}
    \label{fig:gauss}
\end{figure}

While~\cref{thrm:genESPRIT guarantees} holds for bounded noise,~\cref{thrm:cramer-rao} holds for Gaussian noise.
To empirically connect the two, in~\cref{fig:gauss} we demonstrate the performance of genESPRIT under Gaussian noise, as defined in~\cref{sec:cramer-rao}.
Comparing to~\cref{fig:error scaling} we see that the performance is similar to the shot noise case (using the relation $\sigma = \sqrt{2n^{-1}})$, except that the scaling of the error in the spectrum with $N$ seems to be better for Gaussian noise.
Once again, we observe a significant outlier in the spectrum error at $\sigma=0.1, N=13$.

\begin{table}
    \centering
    \caption{Empirical error scalings under shot noise}
    \begin{tabular}{c||c|c}
        \multirow{2}{.3cm}{$N$} & \multicolumn{2}{c}{$\mu: \ \text{error} \approx \mathcal{O}(L^\mu)$}\\
        \cline{2-3}
                                & $\|\hat G - G\|_F$ & $\|\hat\Omega - \Omega\|_{\text{md},\mathbb{T}}$\\
        \hline
        4 & -0.60 & -1.40\\
        10 & -0.56 & -1.40\\
        20 & -0.55 & -1.31\\
    \end{tabular}\\
    \vspace{.3cm}
    \begin{tabular}{c||c|c}
        \multirow{2}{.2cm}{$n$} & \multicolumn{2}{c}{$\nu: \ \text{error} \approx \mathcal{O}(N^\nu)$}\\
        \cline{2-3}
                                & $\|\hat G - G\|_F$ & $\|\hat\Omega - \Omega\|_{\text{md},\mathbb{T}}$\\
        \hline
        10 & 1.13 & 0.45\\
        100 & 1.27 & 0.45 \\
        1000 & 1.20 & 0.27 \\
    \end{tabular}
    \label{table:scaling exponents}\\
    \vspace{.3cm}
    \begin{tabular}{c||c|c}
        \multirow{2}{.2cm}{$N$} & \multicolumn{2}{c}{$\rho: \ \text{error} \approx \mathcal{O}(\kappa^\rho)$}\\
        \cline{2-3}
                                & $\|\hat G - G\|_F$ & $\|\hat\Omega - \Omega\|_{\text{md},\mathbb{T}}$\\
        \hline
        5 & 0.99 & 1.14\\
        10 & 0.95 & 1.13\\
    \end{tabular}
\end{table}

\begin{table}
    \centering
    \caption{Empirical error scalings under Gaussian noise}
    \begin{tabular}{c||c|c}
        \multirow{2}{.2cm}{$\sigma$} & \multicolumn{2}{c}{$\nu: \ \text{error} \approx \mathcal{O}(N^\nu)$}\\
        \cline{2-3}
                                     & $\|\hat G - G\|_F$ & $\|\hat\Omega - \Omega\|_{\text{md},\mathbb{T}}$\\
        \hline
        $10^{-1}$ &  1.28 & 1.31\\
        $10^{-2}$ &  1.31 & 0.15\\
        $10^{-3}$ &  1.61 & 0.19\\
    \end{tabular}    
    \label{table:scaling exponents gauss}
\end{table}

\section{Discussion}\label{sec:discussion}
\noindent
For estimating the generator in 
Frobenius norm of the corresponding propagator, the guarantees~\cref{eq:generator guarantee,eq:generator error} match the scaling in $L$, $N$ and $\sigma$ of the lower bound from~\cref{cor:gaussian estimator} up to logarithmic factors.
The scaling is confirmed by the numerical experiments, see~\cref{table:scaling exponents}.
This indicates that the bound in~\cref{cor:gaussian estimator} is the actual spectrum-independent Cramér--Rao bound, even though the actual Cramér--Rao bound could be better for degenerate spectra.

In~\cref{ssec:generator guarantees} we have translated the error in the propagator to an error in the generator in the case when we have $\|\Omega\|_\infty \le \frac\pi2$.
For this we had to include an extra step of projecting $\mathcal{A}$ to the unitary group into genESPRIT.
We suspect this is not necessary as confirmed by the numerics~\cref{sec:numerics}.
To be able to remove this inconvenience, techniques other than~\cite{kittanehLipschitzFunctionsNormal1985} for Lipschitzness of matrix functions would have to be used.

Considering the scaling of the precision of frequency estimation in $L$, the lower bound in~\cref{thrm:cramer-rao} allows the scaling $L^{-\frac32}$, which is a factor of $\sqrt{L}$ better compared to the guarantee (up to logarithmic factors) and which matches the single-tone frequency Cramér--Rao bound from ref.~\cite{rifeSingleToneParameter1974}.
But this bound comes from the Cramér--Rao bound on the simpler problem of estimating $\Omega$ given knowledge of $A$ and $B$.
It is therefore not clear whether it is possible to achieve this scaling for generator estimation.
For comparison, the $L^{-1}$ scaling in the guarantee matches the best possible worst-case precision (the so-called minimax rate) of estimating well-separated frequencies of a partially clustered spectrum from scalar data~\cite{batenkovSuperresolutionNearcollidingPoint2021}.
This scaling is also shown by atomic norm and BLASSO algorithms~\cite{azaisSpikeDetectionInaccurate2015,tangMinimaxLineSpectral2015,Fern1307Support}.
On the other hand, an algorithm matching the single-tone Cramér--Rao bound has recently been found in the fully separated case~\cite{liApproximateSupportRecovery2020}.
Numerically, we observe for genESPRIT best-fit exponents between $-1.40$ and $-1.31$.

The scaling in $N$ of the error in the genESPRIT frequency output $\hat\Omega$ falls short of the lower bound on the Cramér--Rao bound both analytically and numerically.
First of all, the matching distance bound~\cref{eq:freq guarantee md} has an extra factor of $N$ compared to the frequency-by-frequency bound~\cref{eq:freq guarantee} as required by~\cite{sunVariationSpectrumNormal1996}, but this extra factor is never observed in the numerics.
We therefore conjecture that it can be removed.
The frequency-by-frequency bound~\cref{eq:freq guarantee} gives at first order in the noise a $\sqrt{N}$ scaling, while the lower bound on the Cramér--Rao bound~\cref{eq:freq cramer-rao} is independent of $N$.
The $\sqrt{N}$ scaling is also observed numerically.
At second order in the noise~\cref{eq:freq guarantee} gives an $N^2$ scaling, which we do not see in the numerics, even though this term is leading in terms of the scaling in $L$.
It would be interesting to tighten the scaling in $N$ of the genESPRIT guarantee, to find algorithms that improve upon this scaling, and tighten the bounds on the true Cramér--Rao scaling.

Even though we have restricted ourselves to unitary $A$ and $B$ for computing the guarantee, genESPRIT is capable of dealing with the case of general invertible $A$ and $B$.
As numerically demonstrated, the performance is then controlled by the condition numbers $\kappa(A)$ and $\kappa(B)$.
GenESPRIT in principle also works for complex spectra $\Omega$.
However, in this case it is likely necessary to modify the algorithm somewhat to achieve optimal performance, to accommodate for the potential exponential decay or blow-up of the eigenvalues as $L$ increases.
A more precise analysis of both of these cases is left to future research.

The uniform bound assumption $\|\eta[\ell]\|_\infty \le \lambda$ on the noise has been used for concreteness and simplicity.
But, concentration inequalities for sums of matrices that are not uniformly bounded (e.g. the matrix Rosenthal inequality) are available~\cite{mackeyMatrixConcentrationInequalities2014} and can be used to replace the matrix Bernstein inequality in~\cref{sec:guarantees} without substantially changing the general proof technique for the guarantees.

\section{Conclusions}\label{sec:conclusions}
We introduce the problem of generator estimation as the task of recovering the generator of a matrix exponential time-series and its spectrum.
For the case of unitary data, We prove non-asymptotic recovery guarantees on the genESPRIT algorithm, which was proposed by the present authors in ref.~\cite{hangleiterRobustlyLearningHamiltonian2024a} in the context of learning quantum Hamiltonians.
The guarantees are independent of the spectral spacing, showing that spectral estimation from a \emph{matrix} time-series overcomes the usual scaling of the accuracy of super-resolution methods with the super-resolution factor.
The guarantees are compared with numerical simulations and rigorous lower bounds on the Cramér--Rao limit of generator estimation.
We find that, up to logarithmic factors, genESPRIT estimates the generator matrix with the spectrum-independent Cramér--Rao scaling $\tilde{\mathcal{O}}\left(\sigma NL^{-\frac12}\right)$ in terms of the effective element-wise noise variance $\sigma$, time-series length $L$ and matrix dimension $N$.
In estimating the spectrum, genESPRIT achieves at first order in the noise the scaling $\tilde{\mathcal{O}}(\sigma \sqrt{N}L^{-\frac32})$ in frequency-by-frequency error, which is, up to logarithmic factors, a factor of $\sqrt{N}$ worse than the lower bound on the Cramér--Rao bound we find.
For finite noise, the scaling becomes $\tilde{\mathcal{O}}(\sigma^2 N^2 L^{-1})$.
The corresponding matching distance guarantees are a factor of $N$ worse.

This work opens up a new approach to the spectral analysis of time-series signals, where matrix structure is used to circumvent the unfavourable scaling of the estimation accuracy with the super-resolution factor.

\section*{Code availability}

The computer code used to generate the plots in this manuscript is available as a Python package at https://gitlab.com/jfuksa/genesprit/.

\section*{Acknowledgements}

Funded by the Deutsche Forschungsgemeinschaft (DFG, German Research Foundation) under Germany´s Excellence Strategy – The Berlin Mathematics Research Center MATH+ (EXC-2046/1, EXC-2046/2, project ID: 390685689), as well as the BMFTR (MUNIQC-Atoms, PraktiQom), the Quantum Flagship (Millenion, PasQuans2), and the European Research Council (DebuQC).

\section*{AI statement}

AI tools have been used for minor improvements of the technical steps and to proof read the manuscript.
The authors take full responsibility for the correctness and the presentation of the work.

\bibliographystyle{IEEEtran}
\bibliography{references}

\onecolumn
\appendix
\crefalias{subsection}{appendix}
\subsection{Cramér--Rao bound}\label{app:cramer-rao}
Consider a probability distribution $p_\lambda(x)$ parametrised by $\lambda\in \mathbb{R}^N$ and the task of estimating $\lambda$ from samples.
The asymptotic limits of this task are given by the Cramér--Rao bound, which states that
\begin{equation}
    \cov(\hat \lambda - \lambda) \ge F_\lambda^{-1}(p_\lambda),
\end{equation}
where $\hat \lambda$ is any estimator of $\lambda$ that uses a single sample from $p_\lambda$, and
\begin{equation}
    \left(F_\lambda(p_\lambda)\right)_{j,k} \coloneqq \mathbb{E}\left[\partial_{\lambda_j} \ln(p_\lambda) \partial_{\lambda_k} \ln(p_\lambda)\right]
\end{equation}
is the Fisher information matrix.
The Fisher information matrix is additive, so if we estimate $\lambda$ from a single sample from the product probability measure $p_\lambda(x_0, \dots, x_L) \coloneqq \prod_{i\in[L]} p^{(i)}_\lambda(x_i)$ we have
\begin{equation}
    \left(F_\lambda(p_\lambda)\right)_{j,k} = \sum_{i \in [L]} \left(F_\lambda(p_\lambda^{(i)})\right)_{j,k}.
\end{equation}

\subsection{Fisher information for generator estimation}\label{app:fisher info details}
\noindent
To find lower bounds on the precision of solving the generator estimation problem, we will simplify the task to estimating $A$ given knowledge of $B$ and $\Omega$ and estimating $\Omega$ given knowledge of $A$ and $B$, obtaining a lower bound on the actual complexity.
We will assume without loss of generality (as explained in the main text) that $B$ is supplied in such a way that its rows in the degenerate subspaces of $\Omega$ are orthonormal.
We will work in the asymptotic limit of low noise and large $L$, such that the Cramér--Rao bound for estimating $\Omega$ on the circle $(-\pi, \pi]$ matches the one for finding it in $\mathbb{R}$.

In the case of generator estimation with i.i.d.\ Gaussian noise, we have $N^2 L$ independent complex random variables with probability distributions
\begin{equation}\label{eq:appendix complex Gaussian pdf}
    p(y_{j,k}[\ell]) = \frac1{\pi\sigma^2} \exp\left[-\frac{\left|y_{j,k}[\ell] - \left(A e^{i\ell\Omega} B\right)_{j,k}\right|^2}{\sigma^2}\right] = \frac1{\pi\sigma^2} \exp\left[-\frac{\left|\eta_{j,k}[\ell]\right|^2}{\sigma^2}\right].
\end{equation}
Therefore, we have that
\begin{equation}
    \mathbb{E}[\eta^2] = \mathbb{E}\left[(\eta^*)^2\right] = 0 \ , \quad\mathbb{E}[\eta \eta^*] = \mathbb{E}[\eta^* \eta] = \sigma^2.
\end{equation}
To compute the Fisher information, we need
\begin{align}
    \partial_{\omega_j} \log{p(y_{r,s}[\ell])} &= \frac{2\ell}{\sigma^2} \Im\left(A_{r,j}^* e^{-i \ell \omega_j} B_{j,s}^* \eta_{r,s}[\ell]\right),\\
    \partial_{\Re(A_{j,k})} \log{p(y_{r,s}[\ell])} &= \frac{2}{\sigma^2} \delta_{j,r} \Re\left( e^{-i\ell \omega_k} B^*_{k,s} \eta_{r,s}[\ell] \right),
    \\
    \partial_{\Im(A_{j,k})} \log{p(y_{r,s}[\ell])} &= \frac{2}{\sigma^2} \delta_{j,r} \Im\left( e^{-i\ell \omega_k} B^*_{k,s} \eta_{r,s}[\ell] \right).
\end{align}
We will also require the finite sums
\begin{equation}
    \mathcal{L}_m(z) \coloneqq \sum_{k=0}^{L-1} k^m z^k,
\end{equation}
where $z \neq 1 \in \mathbb{C}$, which are finite versions of polylogarithms. 
To evaluate these sums, notice that 
\begin{equation}
    \mathcal{L}_m(z) = z \frac{\rm d}{\rm dz} \mathcal{L}_{m-1}(z).
\end{equation}
Since $\mathcal{L}_0(z)$ is a sum of a geometric series, we obtain that
\begin{equation}
    \mathcal{L}_m(z) = \left( z \frac{\rm d}{\rm dz} \right)^m \frac{1-z^L}{1 - z}.
\end{equation}
In particular, this shows that $\mathcal{L}_m(z) \in \mathcal{O}(L^m)$ for $z\in\mathcal{O}(1)$.
Since $y_{r,s}[\ell]$ are independent random variables, through the additive property of the Fisher information we obtain the Fisher information matrices
\begin{align}
\begin{split}
    \left(F_\Omega\right)_{j;k} &\coloneqq \sum_{r,s,\ell} \mathbb{E}\left[\partial_{\omega_j} \ln p(y_{r,s}[\ell]) \partial_{\omega_k} \ln p(y_{r,s}[\ell])\right]\\
                                &= \sum_{r,s,\ell} \frac{\ell^2}{\sigma^2} \left(A^*_{r,j} e^{-i\ell\omega_j}B^*_{j,s} A_{r,k} e^{i\ell \omega_k} B^*_{k,s} + A_{r,j} e^{i\ell\omega_j}B_{j,s} A^*_{r,k} e^{-i\ell\omega_k} B_{k,s}\right) \\
                                &= \frac{2}{\sigma^2} \sum_{\ell\in[L]} \ell^2 \Re\left(e^{i\ell(\omega_j - \omega_k)} (A^\dag A)_{j,k} (BB^\dag)_{k,j}\right)\\
                                &= \begin{cases} 
                                    \frac{L(L-1)(2L-1)}{3\sigma^2} (A^\dag A)_{j,j} (BB^\dag)_{j,j} & j = k\\
                                    0 & \omega_j = \omega_k, \ j\neq k\\
                                    \frac{2}{\sigma^2} \Re\left(\mathcal{L}_2(e^{i(\omega_j-\omega_k)})(A^\dag A)_{j,k} (BB^\dag)_{k,j}\right) & \text{otherwise},
                                \end{cases}
    \end{split}\\
    \begin{split}
        \left(F_{\Re}\right)_{j,k;j',k'} &\coloneqq \sum_{r,s,\ell} \mathbb{E}\left[\partial_{\Re(A_{j,k})} \ln p(y_{r,s}[\ell]) \partial_{\Re(A_{j',k'})} \ln p(y_{r,s}[\ell])\right]\\
                                        &= \sum_{\ell\in[L]} \frac{2}{\sigma^2} \delta_{j,j'} \Re\left((BB^\dag)_{k',k} e^{i\ell(\omega_{k'}-\omega_k)}\right)\\
                                        &= \begin{cases}
                                            \frac{2L}{\sigma^2}\delta_{j,j'} (BB^\dag)_{k,k} & k'=k\\
                                            0 & \omega_{k'} = \omega_k, \ k' \neq k\\
                                            \frac{2}{\sigma^2} \delta_{j,j'}\Re\left((BB^\dag)_{k',k} \mathcal{L}_0(e^{i(\omega_{k'}-\omega_k)})\right) & \text{otherwise},
                                        \end{cases}
    \end{split}\\
    \begin{split}
        \left(F_{\Im}\right)_{j,k;j',k'} &\coloneqq \sum_{r,s,\ell} \mathbb{E}\left[\partial_{\Im(A_{j,k})} \ln p(y_{r,s}[\ell]) \partial_{\Im(A_{j',k'})} \ln p(y_{r,s}[\ell])\right]\\
                                        &= \sum_{\ell\in[L]} \frac{2}{\sigma^2} \delta_{j,j'} \Re\left((BB^\dag)_{k',k} e^{i\ell(\omega_{k'}-\omega_k)}\right)\\
                                        &= \begin{cases}
                                            \frac{2L}{\sigma^2}\delta_{j,j'} (BB^\dag)_{k,k} & k'=k\\
                                            0 & \omega_{k'} = \omega_k, \ k' \neq k\\
                                            \frac{2}{\sigma^2} \delta_{j,j'}\Re\left((BB^\dag)_{k',k} \mathcal{L}_0(e^{i(\omega_{k'}-\omega_k)})\right) & \text{otherwise}
                                        \end{cases}
    \end{split}\\
    \begin{split}
        \left(F_{\Re,\Im}\right)_{j,k;j',k'} &\coloneqq \sum_{r,s,\ell} \mathbb{E}\left[\partial_{\Re(A_{j,k})} \ln p(y_{r,s}[\ell]) \partial_{\Im(A_{j',k'})} \ln p(y_{r,s}[\ell])\right]\\
                                             &= \sum_{\ell\in[L]} \frac{2}{\sigma^2} \delta_{j,j'} \Im\left((BB^\dag)_{k,k'} e^{i\ell(\omega_{k}-\omega_{k'})}\right)\\
                                        &= \begin{cases}
                                            0 & \omega_{k'} = \omega_k\\
                                            \frac{2}{\sigma^2} \delta_{j,j'}\Im\left((BB^\dag)_{k,k'} \mathcal{L}_0(e^{i(\omega_{k}-\omega_{k'})})\right) & \text{otherwise},
                                        \end{cases}
    \end{split}
\end{align}
where
\begin{equation}
    F_A = \begin{bmatrix}
        F_{\Re} & F_{\Re,\Im}\\
        F_{\Re, \Im}^T & F_{\Im}
    \end{bmatrix}.
\end{equation}
The vanishing of the Fisher information in the degenerate subspaces (e.g. $\omega_j = \omega_k, \ j\neq k$) is due to the assumption that the rows of $B$ are orthonormal in the degenerate subspaces of $\Omega$.

\subsection{Proof of \texorpdfstring{\cref{cor:gaussian estimator}}{}}\label{app:gaussian estimator}

In order to prove this statement, we will first evaluate the second and fourth moments of $\|\hat A_0 - A\|_F$ using the facts
\begin{equation}
\begin{split}
    \mathbb{E}\left[\eta_{i,j}[\ell]^*\eta_{k,m}[\ell']\right] &= \sigma^2 \delta_{\ell,\ell'} \delta_{i,k}\delta_{j,m},\\
    \mathbb{E}\left[\eta_{i,j}[\ell]^* \eta_{k,m}[\ell'] \eta_{\alpha,\beta}[\lambda]^* \eta_{\gamma,\delta}[\lambda']\right] &= \sigma^4(\delta_{\ell, \ell'} \delta_{\lambda,\lambda'} \delta_{i,k} \delta_{j,m} \delta_{\alpha,\gamma} \delta_{\beta,\delta} + \delta_{\ell, \lambda'} \delta_{\ell',\lambda} \delta_{i,\gamma} \delta_{j,\delta} \delta_{k, \alpha} \delta_{m,\beta})
\end{split}
\end{equation}
about the random matrices $\eta[\ell]$,
while all the two- and four-point correlators with unbalanced numbers of $\eta$ and $\eta^*$ vanish.
Now, letting $E = \hat A_0 - A = \frac1L \sum_{\ell \in[L]} \eta[\ell]B^\dag e^{-i\Omega\ell}$ we evaluate
\begin{align}
    \mathbb{E}[E_{i,j}E_{\alpha,\beta}^\dag] &= \frac{\sigma^2}{L} \delta_{i,\beta}\delta_{j,\alpha},\\
    \mathbb{E}[E_{\alpha,\beta}E^\dag_{\gamma,\delta} E_{\mu, \nu} E^\dag_{\rho, \sigma}] &= \frac{\sigma^4}{L^2} \left(\delta_{\alpha, \delta}\delta_{\beta,\gamma}\delta_{\mu,\sigma}\delta_{\nu,\rho} + \delta_{\alpha,\sigma}\delta_{\beta,\rho}\delta_{\gamma,\nu}\delta_{\delta,\mu}\right),
\end{align}
since $B$ is unitary.
Similarly to $\eta$, all correlators of $E$ with an unbalanced number of $E$ and $E^*$ vanish.
Now, for any diagonal matrix $D$ with diagonal elements $d_i$ we get
\begin{align}
    \mathbb{E}\left[\|\hat A_0 D \hat A_0^\dag - A D A^\dag\|_F^2\right] &= \mathbb{E}\left[\|EDA^\dag + A D E^\dag + E D E^\dag\|_F^2\right]\\
    \nonumber
    &= \mathbb{E}\left[2 \tr E^\dag E D D^\dag + \tr E^\dag E D E^\dag E D^\dag\right]\\
     \nonumber
    &= \frac{2\sigma^2 N}{L} \tr D D^\dag + \frac{\sigma^4 N}{L^2} \left(N\tr D D^\dag + |\tr D|^2\right),
     \nonumber
\end{align}
since $A$ is unitary.
Putting $D = e^{i\Omega}$ and $D = \Omega$ now shows that
\begin{align}
    \mathbb{E}\left[\|\hat A_0 e^{i\Omega} \hat A_0^\dag - A e^{i\Omega} A^\dag\|_F^2\right] &= \frac{2\sigma^2 N^2}{L} + \frac{\sigma^4 N}{L^2}(N^2 + |\tr e^{i\Omega}|^2),\\
    \mathbb{E}\left[\|\hat A_0 \Omega \hat A_0^\dag - A \Omega A^\dag\|_F^2\right] &= \frac{2\sigma^2N}{L}\|\Omega\|_F^2 + \frac{\sigma^4N}{L^2} (N\|\Omega\|_F^2 + (\tr \Omega)^2),
\end{align}
proving~\cref{cor:gaussian estimator}.

\subsection{Proof of \texorpdfstring{\cref{lemma:u v op norm}}{}}\label{app:u v op norm}

\noindent
We have for any $p\in[L-K+1]$
\begin{align}
    U[p] &= \mathcal{H}_K(y)[p,:] V \Sigma^{-1}\\
         &= \mathcal{H}_K(x)[p,:] V \Sigma^{-1} + \mathcal{H}_K(\eta) V \Sigma^{-1}
         \nonumber \\
         &= U_0[p] \sqrt{P} \id V_0^\dag V \Sigma^{-1} + \mathcal{H}_K(\eta)[p,:] (\id - V_0 V_0^\dag) V\Sigma^{-1} + \mathcal{H}_K(\eta)[p,:] V_0 V_0^\dag V \Sigma^{-1} \nonumber\\
          \nonumber
         &= \sqrt{P} U_0[p] \overline\omega_V \Sigma^{-1} + \mathcal{H}_K(\eta)[p,:] (\id - V_0 V_0^\dag) V\Sigma^{-1} + \mathcal{H}_K(\eta)[p,:] V_0 \overline\omega_V \Sigma^{-1}. \nonumber
\end{align}
Comparing this equality with~\cref{eq:u v ansatz} we obtain
\begin{equation}\label{eq:delta U decomp}
    \delta_U[p] = \sqrt{P} \overline\omega_V \Sigma^{-1} - \overline\omega_U +
    \sqrt{L-K+1} e^{-ip \Omega} A^\dag \mathcal{H}_K(\eta)[p,:] (\id - V_0 V_0^\dag) V \Sigma^{-1} +
    \sqrt{L-K+1} e^{-ip\Omega} A^\dag \mathcal{H}_K(\eta)[p,:] V_0 \overline \omega_V\Sigma^{-1}.
\end{equation}
Now, we will bound all terms.
First, focus on the first two.
Using~\cref{lemma:noise Hankel} we get that with probability $1 -\frac\delta2$ 
\begin{equation}
    f \ge \|\mathcal{H}_K(y) - \mathcal{H}_K(x)\|_\infty
      \ge \|U_0^\dag \mathcal{H}_K(y) V - U_0^\dag \mathcal{H}_K(x) V\|_\infty
      = \|\overline\omega_U \Sigma - \sqrt{P} \overline\omega_V\|_\infty,
\end{equation}
so that using~\cref{thrm:wedin} to get a bound on $\|\Sigma^{-1}\|_\infty$ we get
\begin{equation}
    \|\sqrt{P}\overline \omega_V \Sigma^{-1} - \overline \omega_U\|_\infty \le \|\Sigma^{-1}\|_\infty \|\overline\omega_U \Sigma - \sqrt{P}\overline\omega_V\|_\infty \le \frac{f}{\sqrt{P} - f} \eqqcolon g.
\end{equation}
The third term in~\cref{eq:delta U decomp} can be bounded using~\cref{thrm:wedin} together with~\cref{lemma:noise Hankel} as
\begin{align}
    \|\sqrt{L-K+1} e^{-ip \Omega} A^\dag \mathcal{H}_K(\eta)[p,:] (\id - V_0 V_0^\dag) V \Sigma^{-1}\|_\infty 
    &\le \sqrt{L-K+1} \|\mathcal{H}_K(\eta)\|_\infty \|(\id - V_0 V_0^\dag) V\|_\infty \|\Sigma^{-1}\|_\infty\\
     \nonumber
    &\le \sqrt{L-K+1} f \sqrt{2}g \frac{1}{\sqrt{P} - f}\\
    &\le \sqrt{2(L-K+1)} g^2,
     \nonumber
\end{align}
with probability $1 - \frac\delta2$.
To bound the final term we will need a concentration argument to bound $\|\mathcal{H}_K(\eta)[p,:]V_0\|_\infty$.
Consider
\begin{equation}
    \mathcal{H}_K(\eta)[p,:] V_0 = \frac1{\sqrt{K}}\sum_{q\in[K]} \eta[p+q] B^\dag e^{-iq\Omega}.
\end{equation}
Under the noise assumptions from~\cref{thrm:genESPRIT guarantees}, this is a sum of $K$ independent random matrices with bounded operator norm.
Using the matrix Bernstein inequality (Theorem 1.6.2 in ref.~\cite{troppIntroductionMatrixConcentration2015}) we get that for any $\delta>0$ with probability $1-\frac\delta{2L}$ we have
\begin{equation}
    \|\mathcal{H}(\eta)[p,:]V_0\|_\infty \le \ln \frac{4NL}{\delta} \left(\frac{4\lambda}{3\sqrt{K}} + \sqrt{\frac{16\lambda^2}{9K} + \frac{8N\sigma^2}{\ln \frac{4NL}{\delta}}}\right) = \gamma.
\end{equation}
Using the union bound to condition on the events that $\|\mathcal{H}_K(\eta)\|_\infty \le f$ and that the inequality above holds for all $p\in[L-K+1]$, we get that with probability at least $1 - \delta$ we can bound the final term for all $p\in[L-K+1]$ as
\begin{equation}
    \|\sqrt{L-K+1} e^{-ip\Omega} A^\dag \mathcal{H}_K(\eta)[p,:] V_0 \overline \omega_V\Sigma^{-1}\|_\infty \le \frac\gamma{\sqrt{K}\left(1 - \frac{f}{\sqrt{P}}\right)}.
\end{equation}
Putting everything together, we get the lemma.

\subsection{Deriving generator guarantees from propagator guarantees}\label{app:from exp to generator}

\begin{figure}
    \centering
    \includegraphics[width=.3\textwidth]{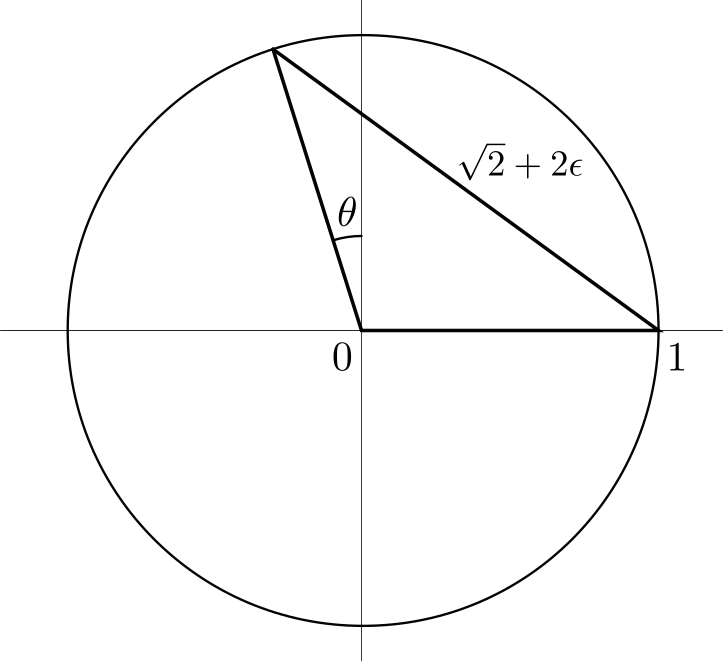}
    \caption{Sketch to bound $\spec(\hat G)$.}
    \label{fig:lip}
\end{figure}

\noindent
Since we know that with probability $1 - \delta$ we have $\|\mathcal{A}' - A e^{i\Omega} A^\dag\|_F \le 2 \epsilon$, from $\|\Omega\|_\infty\le\frac\pi2$ we obtain that
\begin{equation}
    \|\mathcal{A}' - \id\|_\infty \le \|e^{i\Omega} - \id\|_\infty + 2\epsilon \le \sqrt{2} + 2\epsilon.
\end{equation}
Writing the unitary $\mathcal{A}' = e^{i\hat G}$ with Hermitian $\hat G$ satisfying $\|\hat G\|_\infty \le \pi$ (this chooses the logarithm branch), we would like to find $\theta$ such that $\spec(\hat G) \in [-\frac\pi2 - \theta, \frac\pi2 + \theta]$.
\Cref{fig:lip} shows that $\sin(\theta/2 + \pi/4) = 1/\sqrt{2} + \epsilon$, such that
\begin{equation}
    \theta = 2 \left(\sin^{-1}\left(\frac1{\sqrt2} + \epsilon\right) - \sin^{-1} \frac1{\sqrt2}\right) \le \frac{2\epsilon}{\sqrt{1 - \left(\frac1{\sqrt2} + \epsilon\right)^2}},
\end{equation}
using Lipschitz continuity of $\sin^{-1}$ on the interval $\left[\frac1{\sqrt2}, \frac1{\sqrt2}+\epsilon\right]$, 
provided that $\frac1{\sqrt{2}} + \epsilon<1$.

\end{document}